\documentclass[fleqn,usenatbib]{mnras}

\usepackage{newtxtext,newtxmath}

\usepackage[T1]{fontenc}
\usepackage{ae,aecompl}

\usepackage{graphicx}	
\usepackage{amsmath}	
\usepackage{amssymb}	
\usepackage{caption}
\usepackage{graphicx}
\usepackage[utf8x]{inputenc}
\usepackage{lscape}
\usepackage{amsmath}
\usepackage[utf8x]{inputenc}
\usepackage{amssymb}
\usepackage{amsfonts}
\usepackage{tabularx}
\usepackage{longtable}
\usepackage{natbib}
\usepackage{graphicx}
\usepackage{epsfig}
\usepackage{verbatim}
\usepackage[flushleft]{threeparttable}
\include{journaldefs}
\usepackage[caption=false]{subfig}
\usepackage{url}
\usepackage{textcomp}
\usepackage[squaren]{SIunits}
\usepackage{float}
\usepackage[T1]{fontenc} 
\usepackage{aecompl}
\usepackage{adjustbox}
\usepackage{hyperref}
\usepackage{pdflscape}
\usepackage{anyfontsize}

\def\deg{$^{o}\,$}
\def\arcm{$^{\prime}\,$}
\def\arcs{$^{\prime\prime}\,$}
\def\mujybm   {${\rm \mu}$Jy\,beam$^{-1}$}
\def\mjybm   {${\rm m}$Jy\,beam$^{-1}$}
\def\mujy   {${\rm \mu}$Jy}
\def\mjy {${\rm m}$Jy}

\title[HXAGN. I. A radio view at high frequencies]{Hard - X-rays selected Active Galactic Nuclei. I. A radio view at high-frequencies}

\author[E. Chiaraluce et al.]{
\href{http://orcid.org/0000-0002-4090-1327}{E. Chiaraluce,}$^{1,2}$\thanks{elia.chiaraluce@inaf.it}
F. Panessa,$^{1}$
G. Bruni,$^{1}$
R. D. Baldi,$^{3,4,1}$
E. Behar,$^{5}$
F. Vagnetti,$^{2}$
\newauthor
F. Tombesi,$^{2,6,7,8}$
I. McHardy$^{3}$
\smallskip
\\
$^{1}$INAF - Istituto di Astrofisica e Planetologia Spaziali, via del Fosso del Caveliere 100, I-00133 Roma, Italy\\
$^{2}$Dipartimento di Fisica, Univerisità di Roma Tor Vergata, via della Ricerca Scientifica 1, I-00133 Roma, Italy\\
$^{3}$Department of Physics \& Astronomy, University of Southampton, Hampshire SO17 1BJ, Southampton, United Kingdom\\
$^{4}$Dipartimento di Fisica, Universit\'a degli Studi di Torino, via Pietro Giuria 1, 10125 Torino, Italy \\
$^{5}$Physics Department, Technion, Haifa 32000, Israel \\
$^{6}$INAF Astronomical Observatory of Rome, Via Frascati 33, 00078 Monteporzio Catone, Italy \\
$^{7}$Department of Astronomy, University of Maryland, College Park, MD 20742, USA \\
$^{8}$NASA/Goddard Space Flight Center, Code 662, Greenbelt, MD 20771, USA \\
}
\date{Accepted XXX. Received YYY; in original form ZZZ}

\pubyear{2020}
\begin{document}
\label{firstpage}
\pagerange{\pageref{firstpage}--\pageref{lastpage}}
\maketitle

\begin{abstract}

A thorough study of radio emission in Active Galactic Nuclei (AGN) is of fundamental importance to understand the physical mechanisms responsible for the emission and the interplay between accretion and ejection processes. High frequency radio observations can target the nuclear contribution of smaller emitting regions and are less affected by absorption.
We present JVLA 22 and 45 GHz observations of 16 nearby (0.003$\le$z$\le$0.3) hard - X-rays selected AGN at the (sub)-kpc scale with tens \mujybm{} sensitivity. We detected 15/16 sources, with flux densities ranging from hundreds \mujy{} to tens Jy (specific luminosities from $\sim$10$^{20}$ to $\sim$10$^{25}\,W\,Hz^{-1}$ at 22 GHz). All detected sources host a compact core, with 8 being core-dominated at either frequencies, the others exhibiting also extended structures. Spectral indices range from steep to flat/inverted. We interpret this evidence as either due to a core+jet system (6/15), a core accompanied by surrounding star formation (1/15), to a jet  oriented close to the line of sight (3/15), to emission from a corona or the base of a jet (1/15), although there might be degeneracies between different processes. Four sources require more data to shed light on their nature. We conclude that, at these frequencies, extended, optically-thin components are present together with the flat-spectrum core. The ${L_R}/{L_X}\sim10^{-5}$ relation is roughly followed, indicating a possible contribution to radio emission from a hot corona. A weakly significant correlation between radio core (22 and 45 GHz) and X-rays luminosities is discussed in the light of an accretion-ejection framework.  

\end{abstract}

\begin{keywords}
techniques: interferometric - galaxies: active - galaxies: nuclei - galaxies : Seyfert - radio continuum: galaxies - X-rays: galaxies.
\end{keywords}



\section{Introduction}

Active Galactic Nuclei (AGN) show a large variety of radio morphologies, with sources exhibiting compact cores, jets, lobes and knots in radio images, in a wide range of strengths and on scales ranging from sub-pc up to kpc and even Mpc scales. A deep knowledge of the origin of radio emission in AGN is of fundamental importance in order to understand the physics of accretion and ejection onto super-massive black holes (SMBHs), as well as the feedback mechanisms that jets and outflows are thought to produce on the host galaxy. \\
\indent From a radio perspective, AGN are historically divided into Radio Loud (RL) and Radio Quiet (RQ) based on the value of the parameter R=$\log{\frac{f{(4400{\mathring{A}}})}{f{(6\,cm)}}}$, which gives a measure of the strength of radio emission relative to the optical one, with RL objects having R$>$1 and RQ having R$<$1 \citep{Kellerman1989}\footnote{\citet{TerashimaWilson2003} proposed a definition based on the ratio R$_{X}=\log{\frac{{\nu}L_{\nu}(6\,cm)}{L_{2-10\,keV}}}$, with a -4.5 threshold between the two classes.}. Recently, \citet{Padovani2016} proposed a distinction between jetted (with a strong, relativistic jet) and non jetted AGN. However, by adopting this definition caution should be paid as there are examples of relatively low power radio sources with a clear jet like structure, even though not relativistic. For the sake of clarity, we maintain the RL/RQ classification in this work.

While the origin of radio emission in RL AGN, which represent only a minority of the overall population ($\sim$10 per cent), has been identified as synchrotron radiation from relativistic jets \citep[e.g.][]{Begelman1984}, the origin of radio emission in RQ objects is still matter of debate, see \citet{Panessa2019} for a review of the emission mechanisms which may be at work in RQ AGN. There is increasing evidence that RQ are not necessarily radio silent \citep[e.g.][]{Nagar2002}, indeed they emit in the radio, although at lower levels \citep[e.g.][]{HU01}, and they can be associated with outflowing phenomena like jets, but probably less powerful, less collimated and sub-relativistic \citep[e.g.][]{Middelberg2004,GirolettiPanessa2009}. 
However, a complete and systematic census of the radio properties of the AGN population is still missing.

Most of the surveys performed so far comprising RQ objects had suffered from observational biases and were mainly focused on Low-Luminosity AGN (LLAGN), category comprising both low-luminosity Seyferts and LINERs, where radio emission is expected to be ubiquitous \citep{HU01,Nagar2002}. Indeed, these objects would be characterised by a radiatively inefficient accretion flow \citep[as an ADAF, see][]{NarayanYi1994} which would favour the fuelling and launching of collimated outflows/jets, resulting in detectable radio emission; while in Seyferts and quasars at higher accretion rates (but other parameters are thought to be involved, such as black hole mass and spin), the radiative output of the AGN would be dominated by the emission from a standard optically-thick and geometrically-thin accretion disk extending down to the innermost regions of the accretion flow \citep[see review of][]{HeckmanBest2014}.

A tight correlation between the radio and X-rays luminosities has been established for RQ AGN \citep[e.g.][]{Brinkmann2000,Salvato2004,Panessa2007,Panessa2015,Chiaraluce2019}, suggesting a physical connection between the emitting regions.  
The scaling of this relationship with the black hole mass has led to the formulation of the "black hole fundamental plane" where some classes of active black holes have been unified under the same physics \citep{Merloni2003}. 
In particular, LLAGN, similarly to "hard state" X-rays Binaries (XRB), seem to follow an inefficient accretion track \citep{Falcke2004}, while highly accreting AGN, as well as "outliers" hard-state XRBs \citep[e.g.][]{DongWuCao2014}, follow an efficient accretion track \citep[see Figure 7 in][]{Coriat2011}. 
In XRB, the transition between the accretion states (i.e. the hysteresis diagram) is believed to be due not only to the accretion rate \citep[e.g.][]{Maccarone2003}, but also on other fundamental physical quantities, like the disk magnetic field \citep[]{Petrucci2008,BegelmanArmitage2014}, see  \citep[e.g.][]{FenderBelloni2012} for the open issues related to this subject. In AGN there may be as well transitions between one mode and another (see for instance the class of changing look AGN), with some common features with XRB, and still matter of speculation \citep[e.g][]{HeckmanBest2014}.


Extensive studies have been carried out on LLAGN \citep[e.g.][]{HU01,GirolettiPanessa2009,Baldi2018} as well as on more luminous objects, like the PG quasar sample, comprising relatively nearby (z$\le$0.5) high-Eddington objects \citep[e.g.][]{Kellerman1989,Kukula1998}. Interestingly, these works have led to generally high detection rates. 
In both cases, a variety of properties has been reported, with some sources exhibiting compact morphologies, accompanied by jet-like features, which remain compact down to mas scales; the high-brightness temperature and flat-spectrum have led to an interpretation in terms of emission raising from the optically-thick base of a jet; other sources exhibit elongated components, double and triple structures, multiple components as well as rings of radio emission, with spectral indices ranging from steep to flat/inverted \citep[e.g.][]{Kukula1995,BarvainisLonsdaleAntonucci1996,Kellerman1994,Leipski2006,HU01,OrientiPrieto2010,PG13,Baldi2018,Chiaraluce2019}.



This evidence suggests that different physical processes might interplay in RQ AGN, like star formation, both on extended (i.e. few kpc) scale as well as on unresolved (i.e. sub-kpc) scale \citep[e.g.][]{Zakamska2016,Smith2016}, typically associated to a steep GHz spectrum; emission from accretion disc winds in the form of either synchrotron emission from shocks or free-free emission \citep[e.g.][]{BlustinFabian2009,Nims2015}; emission from a compact, flat spectrum opaque synchrotron source interpreted as the base of a jet \citep[e.g.][]{FalckeBiermann1995}.

Since recently, the finding that highly accreting RQ AGN follow the G{\"{u}}del-Benz relation \citep[][]{GudelBenz1993}, i.e. $L_R/L_X\,\sim\,10^{-5}$, valid for coronally active stars, has led to the formulation of coronal models for RQ AGN, in which both radio and X-rays emission would come from a compact corona \citep[e.g.][]{LaorBehar2008}. In this picture, several authors reported the finding of a high frequency (i.e.$\le$95 GHz) excess emission with respect to the extrapolation from the low-frequency (steeper) spectrum \citep[e.g.][]{Baldi2015,Behar2015,DoiInoue2016}. This evidence has been interpreted as signature of the emission from a compact, optically-thick, flat-spectrum component, which dominates in the mm range. This component is recognised to be a corona which is heated through magnetic reconnection events \citep[e.g.][]{RaginskiLaor2016}. 

High-frequency radio observations are particularly useful to target the nuclear contribution of the AGN, eventually testing the emergence of a compact, flat/inverted-spectrum, optically-thick core; indeed radio emission from extended, optically-thin, steep spectrum regions, which may not be directly AGN related, is expected to be resolved out at these frequencies. Moreover, the high frequency approach allows us to characterise the nuclear radio emission of our sources in a frequency regime where smaller regions of emission can be probed due to the improving angular resolution with frequency and in which absorption mechanisms like synchrotron self absorption and free-free absorption are less effective \citep[e.g.][]{Kellermann1966,ParkSohnYi2013}. Only few works have been performed at high frequencies to characterise the radio emission in RQ AGN, one of them being the 1-arcsec, 22 GHz JVLA characterisation of 100 RQ Swift/BAT AGN (0.003$\le$z$\le$0.049) survey performed by \citet{Smith2016,Smith2020}. Various morphologies were found, from compact ($\sim$half of the sample), to extended, interpreted as due to star formation, and jet-like, with flux densities in agreement with both scale-invariant jet models and coronal models. \citet{ParkSohnYi2013} reports single-dish observations for 305 relatively nearby AGN (0.01$\le$z$\le$0.06) at 22 and 43 GHz, obtained from the cross match of Sloan Digital Sky Survey Data Release SDSS-DR7\footnote{Sloan Digital Sky Survey Data Release Seventh} and FIRST\footnote{Faint Image of Radio Sky at Twenty cm} (although the sample contains both LINERs and Seyferts). They found a prevalence of flat/inverted spectra, with however lower detection rates with respect to previous works (i.e. 37 and 22 per cent at 22 and 43 GHz, respectively).

Our strategy is to characterise the high-frequency radio emission of a sample of hard-X-rays selected AGN spanning a wide range of radio-loudness (see right panel of Fig. \ref{fig:kq_histo}) with Jansky Very Large Array (JVLA) observations in C-configuration, which guarantees us a 1-arcsec resolution, that translates in scales of $\le$ kpc. With respect to previous works \citep[e.g.][]{Smith2016,Smith2020}, our novel approach consists in the use of dual frequency observations at 22 and 45 GHz, which allow us to derive high-frequency spectral indices and provide stronger constraints in the discrimination between the diverse radiative processes responsible for the observed radio emission. The hard-X-rays selection of the sample makes it relatively free of selection biases with respect to other wavebands \citep[see discussion in][]{HU01}, giving us the opportunity to have an unbiased view of the radio population. Moreover, our sample of nearby (0.0033$\le$z$\le$0.323) AGN covers, with respect to LLAGN samples and the PG quasar sample, an intermediate range of Eddington ratios ($L_{Bol}/L_{Edd}\,\ge\,10^{-3}$), which enable us to characterise the intermediate accretion regime, for the morphology and energetics study, to compute simultaneous high-frequency spectral indices and test the possible physical scenarios.

This work is part of a larger and comprehensive project aimed at characterising the properties of our hard X-ray selected sample in a wide frequency range, studying the core and extended emission energetics, and the morphology of the radio emission at different frequencies. In a forthcoming work (Chiaraluce et al. in prep), we will couple the high-frequency observations with low-frequency data (C, X and Ku bands, VLA project 19A-018, PI: Chiaraluce). This will enable us to build radio spectral indices and spectral energy distributions (SED), in relation to multi-band data (X-ray, optical, infrared) to derive fundamental parameters on the accretion in light of recent models. Moreover, we have been awarded dual-frequency (3.6 and 6 cm) VLBI observations for 32/44 sources in the sample in a large program of the European VLBI Network (EVN) (code:EC070, 74 hrs). This observations will allow us to spread light on the pc-scale emission of the sources, to break the degeneracy between different physical processes and obtain brightness temperature estimates, useful to discriminate between thermal and non-thermal radiative processes.

In this paper we use a flat $\Lambda$-CDM cosmological model with following parameters: H$_o$=70 km s$^{-1}$ Mpc$^{-1}$, $\Omega_m$=0.3 and $\Omega_{\Lambda}$=0.7 \citep{Jarosik2011}.

\section{Sample}

\begin{table}
\scriptsize
\caption{The 16 sources considered in this work. \textit{Columns:} (1) Target name; (2) $\&$ (3) J2000 positions; (4) Masetti's optical classification. Columns (2)-(5) are from \citet{Malizia2009}.}
\begin{tabular}{ccccc}
\hline
\hline
\multicolumn{1}{c}{Name} & RA  & Dec  & Seyfert & z\\ 
 & (J2000) & (J2000) & Type & \\ 
 (1) & (2) & (3) & (4) & (5) \\ 
 \hline
IGRJ00333+6122 & 00 33 18.41 & +61 27 43.1 & S1.5 & 0.105\\ 
NGC788 & 02 01 06.40 & $-$06 48 56.0 & S2 & 0.0136\\ 
NGC1068 & 02 42 40.71 & $-$00 00 47.8 & S2 & 0.0038 \\ 
QSOB0241+62 & 02 44 40.71 & +62 28 06.5 & S1 & 0.044 \\ 
NGC1142 & 02 55 12.19 & $-$00 11 02.3 & S2 & 0.0288\\ 
B30309+411B & 03 13 01.96 & +41 20 01.2 & S1 & 0.136\\ 
NGC1275 & 03 19 48.16 & +41 30 42.1 & S2 & 0.0175\\ 
LEDA168563 & 04 52 04.85 & +49 32 43.7 & S1 & 0.029 \\ 
4U0517+17 & 05 10 45.51 & +16 29 55.8 & S1.5 & 0.0179\\ 
MCG+08-11-11 & 05 54 53.61 & +46 26 21.6 & S1.5 & 0.0205\\
Mkn3 & 06 15 36.36 & +71 02 15.1 & S2 & 0.0135\\
Mkn6 & 06 52 12.25 & +74 25 37.5 & S1.5 & 0.0188\\
NGC4151 & 12 10 32.58 & +39 24 20.6 & S1.5 & 0.0033\\
NGC4388 & 12 25 46.75 & +12 39 43.5 & S2 & 0.0084\\
NGC5252 & 13 38 16.00 & +04 32 32.5 & S2 & 0.023\\
IGRJ16426+6536 & 16 43 04.70 & +65 32 50.9 & NLS1 & 0.323\\
\hline
\end{tabular}
\label{tab:targets}
\end{table}

The starting point of our work is the sample of 140 extra galactic objects observed by INTEGRAL/\textit{Imager on Board the Integral Satellite} in the 20 -- 40 keV range \citep{Bird2007}. The IBIS survey comprises sources at energies $\ge$20 keV with a sensitivity of $\sim$8$\,\times\,$10$^{-{11}}\,erg\,s^{-1}\,cm^{-2}$. \citet{Malizia2009} applied a Schmitt V/V$_{max}$ test to the IBIS AGN with a significance threshold of $\sim$5 sigma, obtaining a sample of 88 AGN, which comprises 41 Seyfert-1 objects, 5 Narrow-line Seyfert 1 (NLS1), 33 Seyfert-2 and 9 Blazars\footnote{We refer to the optical classification by Masetti, \url{http://www.iasfbo.inaf.it/~masetti/IGR/main.html}, but see also \citet{Masetti2012}}. This sample has several advantages: it is statistically complete and, being hard X-ray selected, it is relatively free of common selection biases affecting samples selected in other ways, like UV-excess and IR \citep[see discussion in][]{HU01}. It has also a wide coverage on information at multi-frequencies, particularly at X-rays, as it has been the subject of several studies \citep[e.g.][]{Panessa2016,Panessa2015,Malizia2014,Molina2013,DeRosa2012,Panessa2011}. Moreover, it represents mainly relatively high luminosity ($41.5\,<\,\log L_{2-10\,keV} (erg s^{-1})<44.5$) and highly accreting ($L_{bol}/L_{Edd}\ge{10^{-3}}$) AGN, with a wide range of X-ray radio loudness R$_{X}$ parameter values (-6.5$\,\le\,R_{X}\,\le\,$-0.5).\\
\indent Considering only the non-blazars objects in the sample (79 out of 88), \citet{Panessa2015} have performed a 1.4 GHz radio characterisation of the 79 AGN with the NRAO Very Large Array Sky Survey \citep[NVSS,][]{NVSS} and the Sydney University Molonglo Sky Survey \citep[SUMSS,][]{SUMSS1,SUMSS2}, having a matched angular resolution of $\sim$45 arcsec and RMS of $\sim$0.45 and 1 mJy beam$^{-1}$ for the NVSS and SUMSS, respectively. They obtained a high detection rate ($\sim$89 percent), with a variety of radio morphologies: unresolved, slightly resolved, on scales from few up to hundreds kpc, exhibiting jet-like features and double/triple structures, as well as diffuse emission. They have also found a significant correlation between the radio luminosity and the X-rays luminosity (either the 2-10 keV or the 20-100 keV one), which have been interpreted as a symptom of an efficient underlying accretion physics, in analogy with 'outliers' hard-state XRBs. \\
\indent In this picture, our group has started a wide-band radio characterisation with the Karl G. Jansky Very Large Array (JVLA) for the part of visible sources in our sample, i.e. north of -30 deg in declination (44 out of 89), with a separate characterisation at high frequencies (i.e. 22 and 45 GHz) and low frequencies (i.e. 5, 10 and 15 Ghz), matched in resolution thanks to the different requested JVLA configurations (C and B, respectively). 

\begin{figure*}
\scriptsize
\centering
\includegraphics[scale=0.275]{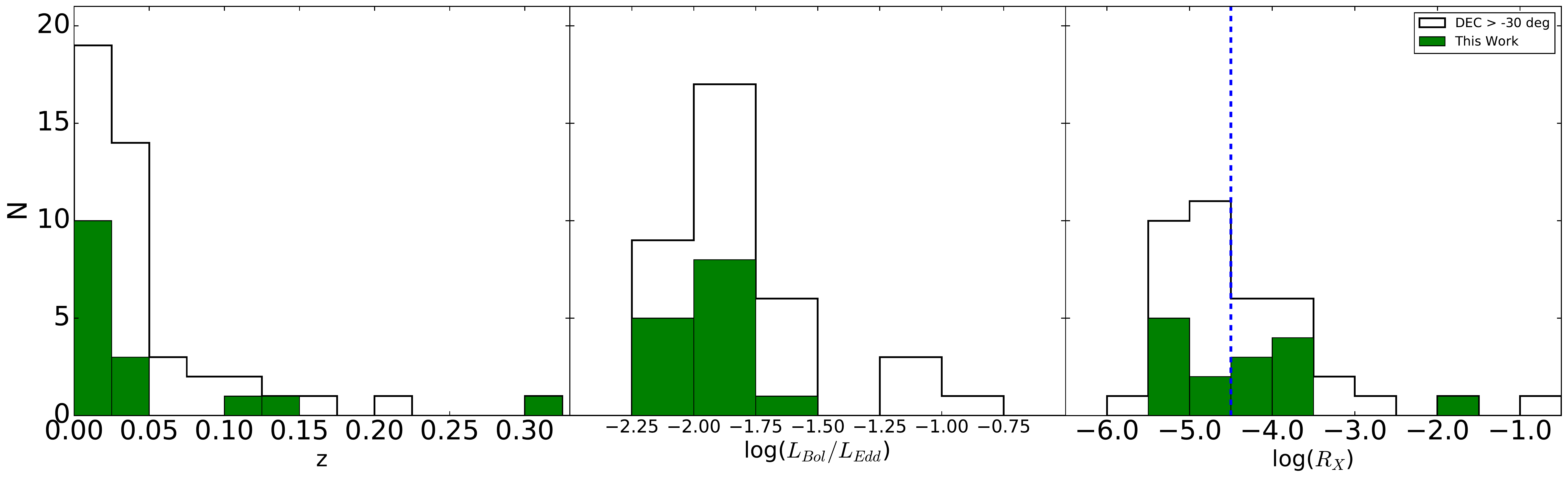}
\caption{Histogram showing the distribution of redshift \citep[from][left panel]{Malizia2009}, Eddington ratio ($L_{Bol}/L_{Edd}$, centre panel) and radio-loudness parameter \citep[from quantities tabulated in][right panel]{Panessa2015} for the parent sample (DEC > -30 deg, semi-filled histogram) and for the sources considered in this work (green). The blue, vertical dashed line is the R$_{X}\,=\,$-4.5 limit of \citet{TerashimaWilson2003}.}
\label{fig:kq_histo}
\end{figure*}

In this work, we report observations for 16 objects out of 44 (see Table \ref{tab:targets}) which have been observed by the JVLA at the high frequencies K and Q (22 and 45 GHz, respectively, VLA project 18B-163) in C configuration (see Table \ref{tab:targets}). The reason for which only 16/44 sources have been observed is because these sources could be observed in the VLA LST range 0 - 12, in which the scheduling pressure (i.e. the number of sessions desiring per hour) is lower\footnote{See  \url{https://science.nrao.edu/observing/proposal-types/tac-reports/vla-pressure-plot-18b-c-updated/view}} . In Fig. \ref{fig:kq_histo} we show the distribution of redshifts, Eddington ratios and radio-loudness parameter for the full sample of 44 objects at DEC > -30 deg (which we will call the 'Parent' sample) and for the objects considered in this work ('Reference' sample). A Kolmogorov-Smirnov test performed considering the three aforementioned quantities to compare the parent and the reference sample results, in all the three cases, in a high (P > 0.7) p-value, suggesting that the null hypothesis that the two samples are drawn from the same parent population cannot be rejected, suggesting that the Reference sample represents the physical properties of the Parent sample. 

Considering the radio loudness of our sources, we have computed the optical radio loudness parameter $R$ for the 16 sources in our sample, either using archival 5 GHz flux measurements, when available, or extrapolating the flux density at 5 GHz from the 22 GHz one, with the 22$-$45 GHz spectral index from our observations (see following sections). We have found that 8 sources can be classified as RL, i.e. IGRJ00333+6122, NGC1068, QSOB0241+62, B30309+411B, NGC1275, LEDA168563, MCG+08-11-11 and Mkn6. 
However, we caution that a distinction between RL and RQ objects based on the value of the radio-loudness parameter may result in an incorrect and incomplete picture about the strength of radio emission relatively to other bands, both because of systematics, as the B flux being dominated by host galaxy in obscured type 2 objects \citep[e.g.][]{Padovani2016} and, as noted before, because it may not take into account ejection phenomena of various morphologies and strengths, which have been observed in objects traditionally classified as RQ.

\section{Observations and data reduction}

A total of $\sim$2.2 hrs of Time-On-Source has been collected across 4 observing runs in the period November - December 2018. In Table \ref{tab:radObs} we list our targets, together with the observing date and the calibrators, grouped by observing block.

Each source has been observed at two frequencies: 22 and 44 GHz, corresponding to the K and Q bands. The observing bandwidths for the two frequencies were 8 GHz, and all bands were subdivided into 64 spectral windows (spw) of 128 MHz bandwidth, and each spw were subdivided into 64 channels.
In Table \ref{tab:radObs} we indicate the sources with corresponding Time-On-Source at each band and the theoretical RMS achievable. The observation of each source has been bracketed between the observations of a nearby phase calibrator, chosen as the nearest one in a radius of 10 degrees, for at least a minute. At the high-frequencies the main responsible for phase fluctuations is the troposphere, and when the Time-On-Source was too large to exceed the recommended cycle time (which ensures a good phase-tracking), we decided to split it into two or more scans on the source, in order to perform a good phase calibration. The absolute flux density calibrator, for each group, has been observed for a total of 7m 30s at the beginning or ending of the block.

In order to reduce and calibrate our data we used the Common Astronomy Software Application \citep[\textsc{casa} 5.4.1 version\footnote{\url{https://casa.nrao.edu}},][]{McMullin+07}. The full raw datasets were downloaded from the NRAO science data archive\footnote{https://archive.nrao.edu/archive/advquery.jsp} as SDM-BDF datasets with flags generated during the observations applied. The calibration has been performed via the CASA pipeline of the same version. After calibration, the plots were inspected for residual RFI and the two bands were split into separate ms files.

Initially, the imaging strategy for the two frequencies has been the following. We considered the deconvolution algorithm of \citet{Hogbom+1974} in \textsc{tclean} with image size of 1024 pixels, with cell sizes of 0.14 and 0.08 arcsec per pixel for the K and Q bands, respectively, which correspond to roughly 1/6 of the beam size for both frequencies. In the two frequency bands the initial weighting algorithm used has been the \citet{Briggs1995} algorithm with robustness parameter equal to 0.5, which ensures a balance between resolution and sensitivity. However, in order to correctly derive the spectral index between the K and Q bands, we also computed tapered radio maps at the Q band, in order to have a matched resolution, with a natural weighting algorithm, in order to give more weight to the short baselines. In this way the observations have an approximately equal UV coverage in wavelengths, and the spectral index is calculated by comparing flux densities obtained from the same emitting regions: this is important for sources exhibiting diffuse, resolved morphology, as it may otherwise led to an artificial steepening of the spectral index. 

For sources presenting imaging artifacts, we first inspected the calibrated visibilities plots in search of unflagged RFI signatures, which we eventually manually flagged. Then, in order to remove the artifacts and enhance the signal-to-noise ratio of the maps we performed a self-calibration, as our sources are bright enough to allow such a procedure. Typically, two cycles of phase-only self calibration were enough to remove the artifacts, but in some cases the two phase-only steps were followed by a phase-amplitude self calibration, which led to the final maps.

For the phase calibrators and absolute flux density calibrators, the same imaging technique as before has been applied. In the case of flux calibrators, we measured the integrated flux densities via a gaussian elliptical fitting on the image plane as performed by the \textsc{casa} \textsc{imfit} and we compared the results with modelled flux densities quoted by \citet{PerleyButler2017}. All the measured flux densities have been found to agree within the 5 per cent with tabulated flux densities, so we adopt an average 5 per cent flux calibration error. The positional accuracy of the phase calibrators is order of few mas, therefore the estimates of radio positions are dominated by uncertainty in the determination of the peak of the gaussian fit as performed by \textsc{imfit}.
The positions, peak intensities, integrated flux densities, deconvolved sizes and position angle (PA) of the sources were estimated by fitting a two-dimensional single or multiple Gaussian in the image plane via the \textsc{casa} task \textsc{imfit}. We determined the rms noise of each map from a source-free annular region around the source.
The uncertainties in the final flux density measurements are affected by fitting errors from \textsc{imfit}, and flux calibration error of 5 per cent, which are added in quadrature and adopted as the error measurements. The positional accuracy of the detected radio components is limited by the positional accuracy of the phase calibrators, typically few mas, and by the accuracy of the gaussian fit to the source brightness distribution as performed by \textsc{imfit}. Therefore, the uncertainty on radio position expressed in this work is the sum in quadrature of the two contributions.
In cases in which the morphology is resolved and in which a multiple gaussian fit is not appropriate, the parameters associated to the emitting components have been estimated using interactively defined boundaries via the \textsc{casa} task \textsc{viewer}. The uncertainties associated to the peak and integrated flux densities are given by the formula ${\sigma_S}=\sqrt{N\times{(rms)^2}+(0.05{\times}S)^2}$, where N is the number of beam areas covered by a source of flux density S, and it is taken into account an uncertainty of 5 per cent in the absolute flux density scale \citep[see][]{HU01}. \\
\indent In Table \ref{table:fluxTable} we summarise our imaging results, and in Figs. \ref{fig:Mappe1} - \ref{fig:Mappe4} we show the coloured and contour maps of the detected sources. 


\begin{table*}\footnotesize
\centering
\caption{List of calibrators (flux and phase) per observation group. \textit{Columns:} (1) Target name; (2) Observation date; (3) Absolute flux density scale calibrator; (4) Phase Calibrator; (5) Observing time in K band for the science target; (6) expected (theoretical) RMS in K band; (7) Observing time in Q band for the science target; (8) expected (theoretical) RMS in Q band.}
\footnotesize
\begin{tabular}{cccccccc}
\hline
 & & \multicolumn{2}{c}{Calibrators} & & & & \\
\cline{3-4} \\
Target & Obs Date & Flux & Phase &  Time & $\sigma_{\mathrm{th}}$ & Time & $\sigma_{\mathrm{th}}$  \\ 
  &  (dd/mm/yy) &  & & (s) & (\mujybm) & (s) & (\mujybm) \\ 
 (1) & (2) & (3) & (4) & (5) & (6) & (7) & (8) \\ 
 \hline
 \hline
NGC788 & 21/11/18 & 3C138 & J0209-0438 & 90 & 25 & 300 & 39\\ 
NGC1068 & 21/11/18 & 3C138 & J0217+0144 & 90 & 25 & 320 & 38 \\
NGC1142 & 21/11/18 & 3C138 & J0312+0133 & 898 & 8 & 300 & 39 \\
4U0517+17 & 21/11/18 & 3C138 & J0510+1800 & 90 & 25 & 300 & 39 \\
\hline 
IGRJ00333+6122 & 21/11/18 & 3C147 & J0109+6133 & 125 & 26 & 310 & 39 \\ 
B30309+411B & 21/11/18 & 3C147 & Self-cal & 95 & 25 & 125 & 61 \\ 
LEDA168563 & 21/11/18 & 3C147 & J0533+4822 & 140 & 21 & 349 & 36 \\ 
\hline 
NGC4151 & 21/11/118 & 3C286 & J1206+3941 & 90 & 25 & 300 & 39 \\ 
IGRJ16426+6536 & 21/11/118 & 3C286 & J1645+6330 & 858 & 8 & 300 & 39 \\ 
\hline 
QSOB0241+62 & 02/12/18 & 3C147 & Self-cal & 60 & 30 & 90 & 72 \\ 
NGC1275 & 02/12/18 & 3C147 & Self-cal & 60 & 30 & 90 & 72 \\ 
\hline 
MCG+08-11-11 & 16/12/18 & 3C147 & J0607+4739 & 90 & 25 & 329 & 37 \\ 
Mkn3 & 16/12/18 & 3C147 & J0524+7034 & 150 & 20 & 170 & 52 \\
Mkn6 & 16/12/18 & 3C147 & J0714+7408 & 90 & 25 & 300 & 39 \\
\hline
NGC4388 & 21/12/18 & 3C286 & J1218+1105 & 900 & 8 & 300 & 39 \\ 
NGC5252 & 21/12/18 & 3C286 & J1405+0415 & 90 & 25 & 120 & 62 \\ 
\hline
\end{tabular}
\label{tab:radObs}
\end{table*}

\section{Results}




\subsection{Detection rates}

We define a source as detected if the peak intensity is above $\ge$5 $\sigma$, where $\sigma$ is the image RMS. However, a source exhibiting a peak intensity 3$\le\,S_P\,\le$5 $\sigma$ is defined as marginal detection, following the same criterion in \citet{HU01}. 

Our strategy results in a high detection rate, with 15 out of 16 sources detected,
which results in a detection rate of 89$\pm$16 per cent~\footnote{Given the small statistics of our sample, a meaningful way of giving a detection rate is via the Laplace point estimate formula (Laplace 1812), which in our case give us a detection rate of 89 per cent for the full sample, with an associated uncertainty which can be estimated via the Adjusted Wald method \citep[e.g.][]{SauroLewis2005} as half the Confidence Intervals.}. The majority of the sources are detected with a high significance (S$\gg$10$\sigma$), only two sources, namely NGC~788 and IGR~J00333+6122, are detected with a significance in between 5$\sigma$ and 10$\sigma$.


The detection rate obtained for our hard-X-rays selected sample is higher with respect to results obtained at lower frequencies for both LLAGN \citep[e.g.][]{HU01} and brighter RQ AGN \citep[e.g.][]{Kellerman1989,Kukula1998}, also at higher frequencies. Indeed, in the mm-range (i.e. $\nu\,\ge\,$95 GHz) \citet{Behar2018} found a detection rate of $\sim$77 per cent for a sample of highly-accreting AGN \citep[see also][for the LLAGN in the mm range]{Doi2011}. Our result confirms a trend which have been observed at these frequecies by \citet{Smith2020} for 100 low-redshift AGN selected at 14-195 keV from the Swift-BAT, 96 per cent, however it is not in agreement with the lower detection rates found by \citet{ParkSohnYi2013} with single dish measurements at 22 and 43 GHz (37 and 22 per cent, respectively).


One source in our sample, i.e. IGR~J16426+6536, has not been detected, and we provide 3-sigma upper limits of 24 and 162 ${\mu}$Jy beam$^{-1}$ at K and Q bands, respectively, corresponding to radio powers of $\sim$6$\times$10$^{20}$ and $\sim$3$\times$10$^{21}\,W\,Hz^{-1}$) at the two frequencies. This source is classified as Narrow Line Seyfert 1 (NLS1) and was neither detected in the NVSS survey \citep[][upper limit on peak intensity of 1.2 mJy]{Panessa2015}. 

\subsection{Flux densities and luminosities}

The radio flux densities of the detected sources (indicated in Table \ref{table:fluxTable}), range from hundreds of $\mu$Jy for the faintest ones, i.e. IGR~J00333+6122 and NGC~788, to tens of mJy, which translates into radio luminosities in the range of 37$\le\,log{{\nu}L_{\nu}(erg s^{-1})}\,\le$40. \\
\indent Considering RL sources, three of them exhibit significantly higher radio flux densities at both frequencies with respect to all the other sources (see Fig. \ref{fig:radiox}). \\ 
\indent For these three sources, the measured flux densities range from hundreds of mJy (QSO~B0241+62), to few Jy (B3~0309+411B) up to tens of Jy (NGC~1275). These sources are known to be powerful radio sources \citep[e.g.][]{Lister1994,Healey2007,Bruni2019}, and a simple comparison with previous works is not straightforward as, besides the variety of resolutions, sensitivities and frequency bands used, the flux densities in literature may be affected by radio variability. Indeed, B3~0309+411 is known for having a variable core flux density \citep[e.g.][at 1.4 and 5 GHz]{Konar2004} and NGC~1275 is an example of recurrent jet activity \citep[e.g.][]{Nagai2010}, therefore our flux densities embed the variable contribution of multiple sub-kpc components. 
For the rest of RL (5/8) and RQ sources there are no significant differences in the flux densities, which range from hundreds of \mujy{} up to tens of \mjy{}, with some extended components of resolved sources up to few hundreds of \mjy{}. \\
\indent These flux densities are in agreement with previous surveys performed at comparable resolution and sensitivities with the VLA, although most at lower frequencies \citep[e.g.][]{Kellerman1989,BarvainisLonsdaleAntonucci1996,Kukula1998,Leipski2006}, and with works performed in the same frequency coverage and with samples with similar characteristics, as \citet{ParkSohnYi2013} and \citet{Smith2020}. 
Our flux densities are also in agreement with 3-mm CARMA flux densities derived by \citet{Behar2018} for 26 hard-X-rays selected AGN from Swift-BAT. If we consider NGC~4388, then the flux of the core component, 3.6$\pm$0.3 mJy, is compatible with our estimate at 22 GHz.


\subsection{Morphology and spectral indices}

Our dual frequency approach at high frequency allows us to take a step forward in the state-of-the-art and compute high-frequency spectral indices and test the occurrence of compact versus jet-like features in a sample of hard-X-rays selected AGN.

We define a spectral index as $S_{\nu}^I\,\propto\,\nu^{-\alpha}$, where $S_{\nu}^I$ is the integrated flux density. In Table \ref{table:fluxTable} we report the spectral indices for all the detected sources and components, in which the spectral index has been calculated considering the integrated flux density measured in K - band maps and in the naturally-weighted, tapered Q - band maps, in order to have matched resolutions. The uncertainties associated to the spectral indices has been estimated as $\sqrt{{{(\sigma_{f_1}/S_{f_1})}^2}+{{(\sigma_{f_2}/S_{f_2})^2}}}/ln({f_2/f_1})$, where $\sigma_{f_{1,2}}$ and $S_{f_{1,2}}$ are the uncertainties on the flux density and the flux density at the two frequencies \citep{HU01}, which are the central frequencies of the K and Q bands (therefore our flux densities are the mean across the bandwidth).

Considering maps in Figs. \ref{fig:Mappe1} - \ref{fig:Mappe4} and deconvolved sizes in Column (7) of Table \ref{table:fluxTable}, then all sources have a compact core component.

No extended emission is detected in 8 out of 16 sources that show a core dominated morphology. Considering the redshift of the sources, this means that they are compact on linear scales smaller than, on average, $\sim$ 1.5 kpc, i.e smaller than 2.5 kpc for the highest-redshift source, B30309+411B, while as low as $\sim$ 70 pc for the lowest redshift one, NGC~4151. For most of our sources the spatial scales sampled by our sources ranges from hundreds of pc to tens of pc, while for two sources, namely B30309+411B and IGRJ00333+6122, the observations maps scales of order of $\sim$2 kpc. We will discuss theses two cases in Section 5. The spectral indices for the compact sources range from flat/inverted to steep\footnote{We define a steep spectrum as having $\alpha\,\ge$0.5, and a flat one as having $\alpha$<0.5, as in \citet{PG13}. We define a spectrum as inverted if $\alpha$<0.}, with a prevalence of steep spectra, as also found in previous VLA studies \citep[e.g.][although at lower frequencies]{Kukula1998}. 

The remaining 7 sources are characterised by a variety of morphologies.
Three sources, namely NGC~4151, Mkn~6 and MCG~+08-11-11, exhibit an elongated morphology, in the direction NE-SW, N-S and N-S, respectively. In the case of Mkn~6, the double morphology seen in the Q band map is not visible in the naturally-weighted, tapered Q-band map. Two sources, i.e. Mkn~3 and NGC~4388, display a double morphology, extending on linear scales of 810 pc (NGC~4388) and 1.1 kpc (Mkn~3) in the direction NE-SW and E-W, respectively. Two sources, NGC~1068 and NGC~1142, exhibit a complex morphology, in which multiple emitting components are visible, extending on linear scales of $\sim$ 1.2 and $\sim$ 11 kpc, respectively. In Table \ref{table:fluxTable}, for clarity, we tabulated first the information for the flux of compact sources, and then that for the resolved sources, indicating the single components.

When considering the radio loudness, no bimodality is found in the observed morphologies and spectral indices, as both RL and RQ objects tend to have both steep and flat/inverted spectra. 


\subsection{Radio-loudness and Eddington ratios}

\begin{figure}\scriptsize
\centering
\includegraphics[scale=0.57]{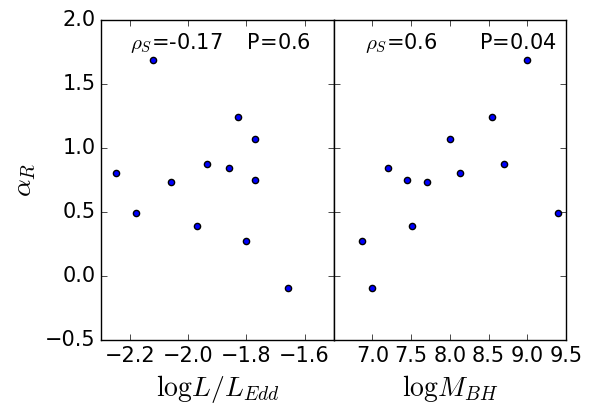}
\caption{Core spectral index as a function of Eddington ratio (left panel) and BH mass (right panel) in which NGC1275, B3 0309+411B and QSOB0241+62 have not been considered. The results of a Spearman's correlation coefficient analysis are reported in the top.}
\label{fig:alpha_edd_mbh}
\end{figure}

Adopting the optical radio loudness parameter, 8/16 sources result as RL AGN. However, this definition may be affected by several systematic uncertainties, for example in the case of type-2 objects, in which the B-band flux may be dominated by host galaxy emission \citep[e.g.][]{Padovani2016}. For this reason, we also computed the radio loudness parameter in the \citet{TerashimaWilson2003} definition, i.e. $R_X\,=\,{L_R}/{L_{2-10\,keV}}$, in which, instead of the 5 GHz luminosity, we used our 22 and 45 GHz core luminosities, which allow us to compare the nuclear radio emission of the cores with the nuclear X-ray luminosity. According to this, 7 sources are classified as RL, i.e. NGC1068, QSOB0241+62, B30309+411B, NGC1275, MCG+08-11-11, Mkn6 at both frequencies, NGC5252 RL only at 22 GHz, 4U0517+17 RL only at 45 GHz. In \citet[][]{Panessa2015}, a continuous distribution of values is found, with a prevalence of RQ objects. A continuous distribution of radio-loudness is found as well by \citet{Burlon2013} at 20 GHz with ATCA and BAT luminosities (although at higher values, -4$\le$R$_{X}\,\le$0). From the hard X-ray selection, the resulting fraction of RL AGN is therefore higher than the 10 per cent usually found from low frequency observations \citep[e.g.][]{Panessa2019}. It is clear that the radio loudness parameter highly varies depending on the adopted definition, therefore should not be taken as a characterising parameter for AGN.
Three sources are occupying a different locus in the radio (either 22 or 45 GHz) X-rays plane with respect to the rest of sources (which comprise both RL and RQ objects, in either of the definitions). Indeed, they are know powerful radio sources and exhibit compact cores and flat/inverted spectra, which can be interpreted as due to a jet closely aligned to the line of sight. \\
\indent We can investigate the relation between the derived physical properties of our sample, i.e. radio 22 and 45 GHz core peak luminosities and spectral index, and relevant physical quantities of the black hole like the Eddington ratio and the black-hole mass, indicated in Table \ref{table:Tab}. In order to calculate the Eddington ratio, we used the recipe in \citet{Lusso2012} (see their Table 3), in particular the formula $L/L_{Edd}\,=\,q/[(L_{X}/L_{Edd})^{-1}-m]$, where $m$ and $q$ are two parameters which have different values depending on whether the object is a type 1 or 2. The X-ray luminosities are from \citet{Panessa2015}, and are absorption corrected. In order to calculate the Eddington luminosities, we have compiled BH masses from literature. The most reliable methods for the determination of the black hole mass are via reverberation mapping measurements \citep[e.g.][]{Peterson2003,McLureJarvis2002}, and by considering the spatially resolved kinematics of stars or gas in close proximity to the black hole \citep[e.g.][]{FerrareseFord2005,YongWebsterKing2016}, but both methods are limited by the statistics of objects studied. Another method which results in the most precise estimates of the black hole mass is via masers measurements \citep[e.g.][]{Greene2016}. Only in four sources, namely NGC4388, Mrk3, NGC1275 and NGC1068, maser emission has been detected, but only for NGC4388 and NGC1068 mass estimated from masers are available, and we report them in Table \ref{table:Tab}. When masses from above measurements were not available, values calculated from other methods have been considered, as using the relations between the black hole mass and the stellar velocity dispersion and bulge luminosity \citep[e.g.][]{FerrareseMerritt2000,Magorrian1998}. Typical uncertainties range from 0.3 dex to 0.7 dex (see References in Table \ref{table:Tab}). \\
\indent We did not find significant trend of our high frequency core spectral indices with either the Eddington ratio or the black hole mass (see Fig. \ref{fig:alpha_edd_mbh}), as instead found by \citet{LaorBaldiBehar2019} for a number of PG quasars at lower frequencies (5 - 8 GHz), in which they isolated the core component considering highest resolution VLA observations. 
They have found a trend according to which high Eddington sources would favour the production of outflows, which would be associated with optically-thin spectral slopes, while in low Eddington ratio object the launching of outflows is quenched and only the compact, flat-spectrum core would be present present, probably of coronal origin. We interpret this discrepancy as first due to a limited statistic, and second to the fact that the emission processes and regions involved at high frequency might be different with respect to lower frequencies. If we compare high frequency spectral slopes with that in literature at lower frequencies at same resolution, we find a general agreement. However, the presence of multiple components found at high frequency makes it difficult unambiguous association of the core, with the effect of a possible dilution of the correlations of the spectral index with either BH mass or Eddington ratio.

\section{Clues on the origin of radio emission}

We presented observations for 16 sources. For 2 of them, namely IGRJ00333+6122 and IGRJ16426+6536, the first radio maps ever have been presented in this work, while for other 3 (NGC1142, LEDA168563 and 4U0517+17), we show the first high-frequency (i.e. 22 and 45 GHz) radio maps.

Our JVLA-C observations, with their $\sim$1 arcsec resolution, allow us to sample the sub-kpc scales for our sources. Indeed, considering only the detected sources, with the exception of B30309+411B and IGRJ00333+6122 (at z$\sim$0.136 and 0.105, respectively), our observations sample spatial scales from few hundreds of pc to tens of pc, and at these scales.

Our sources show a variety of morphologies, with 8 out of 15 being core dominated,
the others being either elongated or exhibiting multiple emitting components. 
The morphology and the spectral index information we obtained at 22 and 45 GHz allow us to try to discriminate between the possible physical mechanisms which may be at work in AGN at these spatial scales, following the flow-chart suggested by \citet{Panessa2019} (see their Fig. 8). In this scheme, if the emission is resolved on arcseconds scales, then a steep spectral index may suggest either star-formation, if the morphology is clumpy and diffuse, or an outflow (or ionisation cone), if symmetric, otherwise a flat component may suggest a free-free emission. However, a compact, flat-spectrum component, unresolved on arcseconds scale, may suggest emission either from the base of a jet or a corona, therefore tracing the position of the core. In this case high-resolution (mas-scale) observations are required. \\
\indent In this picture, we can also consider the additional information about the concentration index $f_C$ of the core component\footnote{The concentration index of the core component has been calculated as $f_C\,=\,{S_{peak}}/{S_{Int}}$, where $S_{peak}$ is the peak intensity and $S_{int}$ is the integrated flux density. For a similar definition see \citet{LaorBaldiBehar2019}.}. The concentration index of the core is a useful tool to quantify the fraction of flux density contained in the compact core. In the case in which in the compact component, identified with the core, additional unresolved components contribute to the emission, then we expect a concentration index <1, as the contribution of these components becomes more and more important; in the case in which the emission of the compact component is actually dominant, a concentration index $\sim$1 is expected. 

In the case of the asymmetric double NGC~4388, the flat-spectrum component (NE) could be associated with the core, with a concentration index $f_{C}\,\ge\,$0.75, while the steeper SW component which could be associated to a jet component. Similar conclusions can be drawn for the other asymmetric double, Mkn~3, in which the core component, having a steeper spectrum, may hide additional (steep) unresolved components ($f_{C}\,\sim\,0.6$). NGC~1068 exhibits a bipolar outflowing structure, with both compact and extended emission, and the interpretation of the spectral indices in this case is not straightforward. Nevertheless, the extended ($\sim$1.2 kpc) structure could be interpreted as the interaction of a radio jet emanating from the nucleus with high-ionisation gas clouds, as also suggested in previous works \citep[e.g.][]{Capetti1997}. NGC~1142, part of the interacting system Arp~118, has a complex morphology, with multiple emitting components distributed on scales of $\sim$19 arcsec in K band, corresponding to $\sim$11 kpc. Component B, identified by the core and having a high ($\ge$0.9) concentration index, is in the centre of a ring-like region, with a southern filament of radio emission and several components at the NE and SW (although only components labelled A1 and A2 are visible at both frequencies). The steep spectrum of the off-nuclear sources, and the complex morphology, suggests a star formation origin of the extended emission, probably produced by the tidal interaction the system is undergoing.

Considering the remaining sources, three of them exhibit an elongated morphology. The maps of NGC~4151 reveal an elongated morphology in the NE-SW direction, with a steep spectrum, compatible with what found at lower frequencies at similar resolution \citep[e.g.][]{Pedlar1993}. This evidence, together with a concentration index of $\sim\,$0.7, could be interpreted as symptom of a non-negligible contribution from optically-thin steep components, from a jet or an outflow \citep[indeed eMERLIN observations at lower frequencies points toward a jet origin, see][]{Williams2017}. Same conclusions can be drawn for MCG~+08-11-11 and Mkn~6, with the elongations due to a low power jet or an outflow.

Finally, 8 out 15 sources are characterised by compact morphologies (either unresolved or resolved). Two sources, IGR~J00333+6122, and LEDA~168563, exhibit steep spectra, but while for the latter the emission mapped by our observations is on scales of order of $\sim$600 pc, in the former they map scales of $\sim$2.5 kpc. In both cases the emission may be due outflowing optically-thin components, with a contribution from unresolved sub-kpc star formation. In the case of IGR~J00333+6122, these steep spectrum components may be on larger scales, i.e. 1-2 kpc, and therefore we can not exclude a contribution from circumnuclear star formation on these scales. For this source, high resolution observations are required to break this degeneracy and disentangle the contribution of different mechanisms, and for this purpose we will benefit of our VLBI observations, although at lower frequencies. NGC~5252 has two components, the southern one having a steep spectrum, the northern one is detected only in K-band \citep[the association of the northern component with the nucleus, the southern component, has been questioned, see][]{Kim2015,YangYangParagi2017}. One source, NGC~788, exhibit a spectrum which is borderline, i.e. compatible between steep and flat, therefore the interpretation is not straightforward. Higher resolution observations are necessary to draw robust conclusions on the nature of radio emission in these sources. \\
\indent We note that a possible contribution to the spectral index of steep spectrum components may come from spectral ageing, as higher energy electrons radiates away their energy first, resulting in a steeping of the slope at high frequencies. However, in order to quantify its contribution, high resolution and high sensitivity observations are required over a wide range of frequencies, at a matched resolution, able to resolve possible sub-kpc structures even for compact sources, and to determine the high-frequency spectral break for various regions of the source, having different electron ages.
Although we can not rule out a contribution from spectral ageing for steep spectrum sources (both resolved and unresolved ones), since our purpose is to use the spectral index as a tool to discriminate between core components and extended, steep spectrum ones, and establish the dominant radiative mechanisms, it does not affect our main considerations.

Four sources show a flat (QSO~B0241+62 and NGC~1275) and inverted (4U~0517+17) spectra and are compact, core-dominated, with B3~0309+411B having a spectrum compatible with both flat and inverted (however for the last source our observations map spatial scales of $\sim$2 kpc). These sources, except for 4U~0517+17, are known powerful radio sources. In these sources, high resolution (i.e. VLBI) monitorings have revealed pc-scale relativistic jets closely aligned to the line-of-sight \citep[MOJAVE Project\footnote{\url{http://www.physics.purdue.edu/astro/MOJAVE/allsources.html}, and references therein for single sources.},][]{Lister2019}. This suggests that the evidence of a compact morphology, a flat spectrum and a enhanced radio flux, due to Doppler boosting, can be interpreted as due to a jet closely aligned with the line-of-sight. Although the observations for B3~0309+411B maps larger spatial scales with respect to the other two sources, the same argument can be applied as well. The source 4U~0517+17 is characterised by a flat spectrum, which can be interpreted as signature of either synchrotron emission from the optically-thick base of a jet or emission from a corona. However, higher resolution observations are required to draw more robust conclusions.


A supplementary appendix with notes on individual sources will be available in the online version of the paper.

\section{The X-ray and radio correlation}

Our high frequency observations allow us to investigate also the existence of a correlation between radio luminosity, at either 22 or 45 GHz, and the X-rays one (i.e. 2$-$10 keV), focusing on the core contribution, as emission from more extended components is expected to fade. Some sources exhibit extended components which may not be AGN related, and some others, despite being compact, may hide extended sources (as suggested by the concentration index), and for this reason we considered the core peak luminosities only. Such correlations have been found (mostly at lower frequencies) for LLAGN and low-hard state XRB, which have been interpreted as symptom of inefficient accretion regime \citep[e.g.][]{PG13}, as well as highly-accreting AGN and 'outliers' XRB, interpreted as symptom of radiatively efficient accretion flows \citep[e.g.][]{Coriat2011,DongWuCao2014}, in a broad picture of common accretion-ejection physics across different BH masses. Excluding the three known powerful sources, we find a correlation of the form $\log{L_R^{peak}}(erg\,s^{-1})\,\propto\,m\,\log{L_{2-10\,keV}}(erg\,s^{-1})$, with $m\,=\,$0.66$\pm$0.26 and 0.56$\pm$0.25 at 22 and 45 GHz, respectively. However, in both cases the statistical significance is low, as a Spearman correlation test results in $\rho\,\sim$0.6 with P$\sim$0.06 at 22 GHz and  $\rho\,\sim$0.5 with P$\sim$0.1 at 45 GHz. No definite conclusion can be drawn from this correlation as more sources are required to strengthen this finding. Indeed, so far the relation has been established at lower frequencies and on different scales \citep[e.g.][]{Panessa2015}, where different processes and emitting regions may be involved.

Finally, we tested whether our sources follow the ${L_{R}}/{L_X}\sim10^{-5}$ relation predicted by coronal models, according to which both radio and X-rays would originate in a magnetically heated corona, the contribution of which should emerge in the (sub)-mm range \citep[e.g.][]{LaorBehar2008,RaginskiLaor2016}. As can be seen from Fig. \ref{fig:radiox}, three powerful radio sources tend to depart significantly from the relation, with the rest of the sources roughly following the relation. This result may be interpreted as an indication of a contribution of a magnetically heated corona to the radio emission at these frequencies. A more robust statistical result will be provided by the new data that will be available on this sample.


\section{Discussion}

The novelty of our strategy with respect to previous works is the use of dual frequency (22 and 45 GHz) observations, which allow us to compute high-frequency spectral indices,  couple them with morphological information and eventually discriminate between possible radiative mechanisms regulating radio emission at the sub-kpc scales, except for two sources, B30309+411B and IGRJ00333+6122, for which our observations map scales of order of $\sim$2 kpc (for the former high resolution observations have been already carried out, for the latter we will benefit of our VLBI observations for the sub-kpc scale characterisation). \\ 
\indent We note that high-frequency radio observations are particularly useful to target the nuclear emission, as at these frequencies (at our resolutions), the contribution of more extended, optically-thin, steep-spectrum components, which may be not AGN-related, is expected to be less and less important, and absorption mechanisms like synchrotron self absorption and free-free absorption are less effective with respect to lower frequencies \citep[e.g.][]{Kellermann1966,ParkSohnYi2013}.

Moreover, the hard - X-rays selection of the sample makes it ideal for an unbiased characterisation of the radio population at these redshifts, which can be noted considering the distribution of flux densities in Table \ref{table:fluxTable}. Our radio characterisation fills a gap in the existing literature. Indeed, our sources are characterised by Eddington ratios at intermediate values between that of LLAGN, usually exhibiting $L_{Bol}/L_{Edd}\,\le\,10^{-3}$, and the PG quasars, usually only type-1 objects characterised by $L_{Bol}/L_{Edd}\,\sim\,0.1-1$, in an intermediate redshift range which partially overlaps with both cases. This gives us the chance to sample the radio emission in an intermediate accretion regime, and test the occurrence of compact vs extended, jet/outflows-like features.

\begin{figure}\scriptsize
\centering
\includegraphics[scale=0.445]{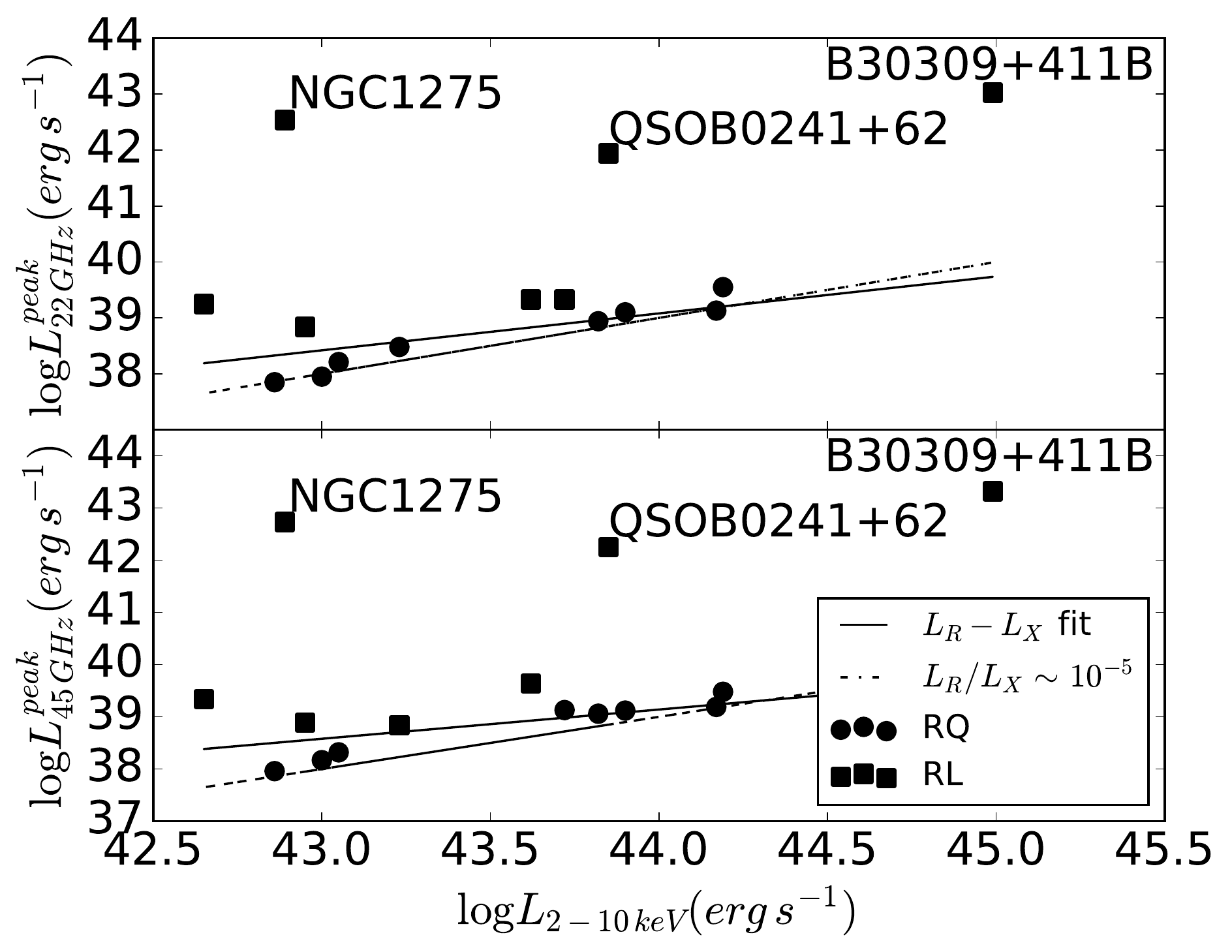}
\caption{The  22 GHz peak luminosity (top panel) and the 45 GHz peak luminosity (bottom panel) of the core components as a function of the 2-10 keV luminosity. Black squares and black circles represents RL and RQ, respectively, considering the \citet{TerashimaWilson2003} with $R_X={L_{22\,GHz}}/{L_{2-10\,keV}}$ (top panel) and $R_X={L_{45\,GHz}}/{L_{2-10\,keV}}$ (bottom panel). The black line is a linear fit performed not considering the three radio galaxies B3 0309+411B, NGC1275 and QSOB0241+62, while the black dah-dotted line is the ${L_R}/{L_X}\,\sim\,10^{-5}$ line.}
\label{fig:radiox}
\end{figure}

Few works have been devoted to the characterisation of the high frequency emission in a general population of AGN, one of them being that of \citet{ParkSohnYi2013}, who characterised the high-frequency (22 and 43 GHz) emission of sample of AGN with single dish measurements. Despite the comparable redshift range and the similar radio luminosities of the detected sources, they find a prevalence of flat/inverted spectra and significantly lower detection rates (37 and 22 per cent at 22 and 43 GHz, respectively). We attribute these differences to a combination of both the different criteria for the selection of the samples and to the single-dish observations. Indeed, the cross match of SDSS-DR7 and FIRST may have led, from a radio point of view, to the selection of powerful objects (because of FIRST sensitivity) with bright extended features which may be not AGN related (because of 5 arcsec FIRST resolution), while single-dish observations may have encompassed radio emission from the AGN core as well as from extended features. \\
\indent More recently, \citet{Smith2016,Smith2020} characterised a sample of hard-X-rays selected AGN from Swift-BAT at 22 GHz at $\sim$1 arcsec resolution, with most of properties similar to our sample, e.g. redshift range, sensitivity of the hard-X-rays instrument, X-ray and hard-X-rays luminosities, and sensitivity of radio observations. The high detection rate found for our observations (15 out of 16 sources) is in agreement with what found by Smith et al., and a similar conclusion can be drawn considering the morphologies of the sources, as they find that more than half of the sources have a compact morphology. However, they also find that a high fraction (30/96 sources) has a morphology compatible with star formation, with only 11 sources compatible with hosting sub-kpc to kpc scale jets. The interpretation of our high-frequency spectral indices and morphologies led us to different conclusions, according to which for only one source, namely NGC1142, the kpc-scale radio emission is star formation related (as a result of tidal interaction in an interacting system), while for other 6 sources it is compatible with presence of a jet. However, this does not exclude the contribution of star formation. Indeed, while our high-frequency, $\sim$1 arcsec resolution observations are effective in isolating the core AGN emission with respect to the extended one, which may be not AGN-related, a contribution from sub-kpc, nuclear star formation may be present as well \citep[e.g.][]{Smith2016,Smith2020}. \\
\indent We have also found, in agreement with \citet{Smith2016,Smith2020}, that our sources, with the exception of three powerful sources (see Fig. \ref{fig:radiox}), roughly follow the G{\"{u}}del-Benz relation ${L_{R}}/{L_X}\sim10^{-5}$ valid for coronally active stars \citep{GudelBenz1993}, which have been interpreted in the sense of a common origin in a magnetically-heated corona of both radio and X-rays emission \citep[e.g.][]{LaorBehar2008}. However, it is not straightforward to disentangle the contribution of a corona from that of a low-power, sub-kpc scale radio jet. Recently, \citet{Baek2019} have performed a VLBI characterisation at 22 GHz for a number of radio bright hard-X-rays selected AGN from the Swift-BAT sample, proving that pc-scale jets in these sources are not common. However, this may not be the case for the fainter hard-X-ray selected AGN (as some members of our sample), in which less powerful, less collimated jets may be present. In our discussion about the origin of the radio emission in our sources we referred to the flow chart in \citet{Panessa2019} (their Figure 8). In less powerful, compact sources, in order to disentangle the contribution of a corona to that of a sub-kpc scale low-power jet, multi-frequency high-resolution observations are required, but the definite proof would be a monitoring of the correlated radio and X-ray variability following the Neupert effect, predicted in coronal models. This promising methodology is only at an early stage \citep[e.g. in][for NGC7469]{Behar2020}. \\
\indent Finally, we note that, although only a part of our sample of 44 sources have been observed (16/44 sources), we expect the properties we have found in this work to be representative of that of the full sample. Indeed, as explained before, the observations of our sources have been carried out in a random way, i.e. without introducing any selection bias, as can be appreciated in histograms in Fig. \ref{fig:kq_histo}. We expect that the future JVLA observations in our program will likely strengthen our findings statistically.

\section{Conclusions}

In this work we presented the results of a high-frequency (22 and 45 GHz) JVLA observational campaign for a sample of relatively nearby (0.0033$\le$z$\le$0.323) hard - X-rays selected AGN. The resolution the JVLA C-configuration guarantees us ($\le$1 arcsec) allow us to probe linear scales of $\sim$70 pc for the nearest one (NGC~4151) and $\sim$2.4 kpc for the farthest one (B3~0309+411B). \\
\indent We can summarise our main results as following: \\
\indent - Our strategy translates into high detection rates at both frequencies, i.e. 15 out of 16, 89$\pm$16 per cent. The only undetected source, namely IGRJ~16426+6536, is a NLS1, for which we derive 3-sigma upper limits at 24 and 162 \mujy{} beam$^{-1}$ level at K and Q bands, respectively. \\
\indent -  The flux densities range from hundreds of \mujy{} up to hundreds of \mjy{} (radio luminosities in the range 37$\le\,log{{\nu}L_{\nu}(erg s^{-1})}\,\le$40).  Three sources, which are known powerful radio sources, have significantly higher flux densities (up to tens of Jy). \\
\indent - We find a compact core in all of the detected sources, of which 8/15 are core-dominated, while 7/15 exhibit extended structures (e.g. double, elongated and multiple components). The spectral indices of the components range from steep to flat/inverted, suggesting that the expected compact, flat-spectrum core may not be dominant and emission from optically-thin, steep spectrum components may be not negligible.\\
\indent - Different physical mechanisms may be responsible for the radio emission, a core + low power jet or an outflow, the base of a jet or a corona , star formation from tidal interaction, or a jet closely aligned to the line-of-sight.  For four sources, the interpretation is not straightforward. \\
\indent - In the case of our sample the radio/X-ray radio-loudness parameter does not separate univocally the RL AGN, which are expected to be powered by relativistic jets, and the RQ AGN where different astrophysical procesess compete (weak jet, SF, disc wind, outflowing magnetically-active corona). We conclude that such a parameter cannot be used to distiguish between the properties of two populations. The three sources well known to be strong radio sources indeed differ significantly from the rest of the sample. 
\begin{table*}
\caption{The derived quantities for the 15 detected objects in the sample. The radio luminosities are that of the core components, when it has been possible to identify them, as well as the radio-loudness parameters R$_X$, and the peak flux density has been considered. Same argument applies for the spectral index. The 2-10 keV absorption-corrected luminosities are from \citet{Panessa2015}, unless otherwise specified.} 
\centering
\begin{adjustbox}{width=1\textwidth,center=\textwidth}
\begin{tabular}{cccccccccccccc}
\hline
Target & L$_{22\,GHz}$ & L$_{45\,GHz}$ & L$_{2-10\,keV}$ & $\alpha$ & $M_{BH}$ & $L/L_{Edd}$ & f$_{C}$(K) & f$_{C}$(Q) & f$_{C}$(Q$_{t}$)& R$_{X}(22\,GHz)$ & R$_{X}(45\,GHz)$ & R$_{O}$ & Ref\\ 
  & (erg s$^{-1}$) & (erg s$^{-1}$) & (erg s$^{-1}$) & & ($M_{\odot}$) &  & &  & & & \\ 
 (1) & (2) & (3)  & (4) & (5) & (6) & (7) & (8) & (9) & (10) & (11) & (12) & (13) & (14) \\ 
 \hline
\hline
 IGRJ00333+6122 & 39.55 & 39.48 & 44.19 & +1.24 & 8.54 & -1.83 & 0.89 & 0.58 & 0.87 & -4.64 & -4.71 & 1.9 & \citet{Masetti2009} \\
 NGC788 & 37.85 & 37.96 & 42.86 & +0.39 & 7.51 & -1.97 & 0.92 & 0.89 & 0.75 & -5.01 & -4.9 & -0.93 & \citet{WooUrry2002} \\
 NGC1068 & 38.838 & 38.89 & 42.95 & +0.84 & 7.2 & -1.86 & 0.77 & 0.57 & 0.76 & -4.11 & -4.06 & 2.11 & \citet{Greenhill1996}\\
 QSOB0241+62 & 41.94 & 42.25 & 43.85$^{*}$ & +0.35 & 8.09 & -1.809 & 0.99 & 0.99 & 0.99 & -1.91 & -1.6 & 1.15 & \citet{Koss2017} \\
 NGC1142 & 38.94 & 39.06 & 43.82 & +0.49 & 9.4 & -2.18 & 0.94 & 0.99 & 0.86 & -4.88 & -4.76 & 0.15 & \citet{Winter2009} \\
 B3 0309+411B & 43.025 & 43.320 & 44.99 & +0.06 & - & - & 0.99 & 0.99 & 0.99 & -1.96 & -1.67 & 3.2 &\\
 NGC1275 & 42.54 & 42.73 & 42.89 & +0.4 & 8.5 & -2.18 & 0.99 & 0.99 & 0.99 & -0.35 & -0.16 & 2.99 & \citet{WooUrry2002} \\
 LEDA168563 & 39.1 & 39.12 & 43.90 & +1.07 & 8.0 & -1.77 & 0.94 & 0.93 & 0.99 & -4.8 & -4.8 & 1.51 & \citet{Vasudevan2009} \\
 4U0517+17 & 38.48 & 38.84 & 43.23 & -0.097 & 7.0 & -1.658 & 0.92 & 0.95 & 0.98 & -4.75 & -4.39 & 0 & \citet{Stalin2011} \\
 MCG+08-11-11 & 39.33 & 39.64 & 43.62$^{*}$ & +0.75 & 7.45 & -1.77 & 0.65 & 0.57 & 0.76 & -4.29 & -3.98 & 1.23 & \citet{Fausnaugh2017}\\
 Mkn3 & 39.13 & 39.19 & 44.17 & +0.87 & 8.7 & -1.936 & 0.60 & 0.47 & 0.61 & -5.04 & -4.98 & 0.95 & \citet{WooUrry2002} \\
 Mkn6 & 39.25 & 39.34 & 42.65$^{*}$ & +0.8 & 8.13 & -2.25 & 0.62 & 0.74 & 0.64 & -3.4 & -3.31 & 1.38 & \citet{Grier2012} \\
 NGC4151 & 38.21 & 38.32 & 43.05 & +0.73 & 7.7 & -2.06 & 0.71 & 0.71 & 0.74 & -4.84 & -4.73 & 0.61 & \citet{Greene2016} \\
 NGC4388 & 37.95 & 38.17 & 43.00 & +0.27 & 6.86 & -1.803 & 0.75 & 0.83 & 0.74 & -5.05 & -4.83 & -1.01 & \citet{Greene2016}\\
 NGC5252 & 39.33 & 39.13 & 43.72 & +1.68 & 9.0 & -2.12 & 0.93 & 0.94 & 0.91 & -4.39 & -4.59 & 0.68 & \citet{Graham2008}\\
\hline
\hline
\multicolumn{14}{l}{Column (1), name of the source; Column (2), 22 GHz peak luminosity; Columns (3), 45 GHz peak luminosity; Column (4), 2-10 keV luminosity; Column (5), core spectral index; Column (6); black hole mass;} \\
\multicolumn{14}{l}{Column (7), Eddington ratio; Column (8), concentration index in K band; Column (9), concentration index in Q band; Column (10), concentration index in tapered Q band;} \\
\multicolumn{14}{l}{ Column(11), R$_{X}(22\,GHz)=\log{L_{22\,GHz}}-\log{L_{2-10\,keV}}$; Column (12), R$_{X}(45GHz)=\log{L_{45\,GHz}}-\log{L_{2-10\,keV}}$; Column (13), R$_{O}=\log{S_{5\,GHz}}-\log{S_{B}}$; Column (14), reference for M$_{BH}$.}\\
\multicolumn{11}{l}{$^{*}$ X-rays luminosity from \citet{Molina2019}.}\\
\end{tabular}
\end{adjustbox}
\label{table:Tab}
\end{table*}

\indent- We found indication of a correlation between the 22 and 45 GHz core luminosities and the 2-10 keV luminosity, with slopes of 0.66$\pm$0.26 and 0.56$\pm$0.25, respectively, suggesting a connection between the accretion and ejection processes holding at relatively higher frequencies. The statistical significance of the relations is low, therefore larger samples are required to derive more robust results. \\
\indent - The sources roughly follow the ${L_R}/{L_X}\,\sim\,10^{-5}$ relation proposed by coronal models \citep[e.g.][]{LaorBehar2008}, suggesting a contribution to the radio emission from a hot corona, with the three powerful radio sources which seems to depart significantly from the relation, as expected.

Our results suggest that even at high-frequencies the compact, flat-spectrum core is present, but it is not always dominant, as originally expected, and that steep-spectrum components on scales mapped by our observations (smaller than $\sim$ kpc) can contribute in a non negligible way.

The high-frequency band in the radio represents a unique tool to study the emission mechanisms of this population of AGN, as it allows to target their nuclear component, thanks to the increasing resolution with frequency, being relatively safe from absorption mechanisms with respect to lower frequencies. A more thorough characterisation at these frequencies will be possible with unprecedented resolution and sensitivity of the next generation interferometers. Indeed, the Square Kilometre Array (SKA) in its phase 1 (SKA1) will extend up to 15 GHz, while in its phase 2 (SKA2) it is expected to extend up to 30 GHz \citep[e.g.][]{Braun2019}, while the next-generation VLA (ngVLA) will cover frequencies up to 116 GHz \citep[][]{Selina2018}.



\section*{Acknowledgements}

EC would like to thank Josh Marvil and Anna Kapinska of the National Radio Astronomy Observatory (NRAO) for the help in the data reduction, and Ari Laor and Daniele Dallacasa for the useful comments to interpret the evidence. The National Radio Astronomy Observatory is a facility of the National Science Foundation operated under cooperative agreement by Associated Universities, Inc.  
EC acknowledges the National Institute of Astrophysics (INAF) and the University of Rome – Tor Vergata for the PhD scholarship in the XXXIII PhD cycle. FP acknowledges support from a grant PRIN-INAF SKA-CTA 2016. 
GB acknowledges financial support under the INTEGRAL ASI-INAF agreement 2013-025-R.1. FT acknowledges support by the Programma per Giovani Ricercatori - anno 2014 Rita Levi Montalcini. This research made use of APLpy, an open-source plotting package for Python \citep{aplpy}.




\bibliographystyle{mnras}
\bibliography{ref} 



\appendix

\section{Notes on individual sources and radio maps}

\smallskip
\noindent\textbf{4U0517+17} 
This source have been studied by \citep{Schmelz1986} with the VLA in C configuration. They put an upper limit of $\sim$1.1 mJy for the flux density at 20 cm. \citet{Panessa2015} derive a flux of $\sim$6.1 mJy with NVSS (45 arcsec resolution), finding a slightly resolved morphology. Our data at 22 and 45 GHz reveal compact morphology, either slightly resolved (at 22 GHz) or unresolved (at 45 GHz). Flux densities are of the order of 2 - 2.5 mJy, and the spectral index is flat/inverted $\alpha\,=$-0.097$\pm$0.11, which can be interpreted as signature of either emission from the optically-thick base of a jet or from a corona.

\smallskip
\noindent\textbf{B3 0309+411B} 
It is a well-known broad line radio galaxy, with Mpc-scale structure, with a milliarcsecond core with a quasar-like spectrum highly varying in the radio \citet{deBruyn1989}. The core is found to have a slightly inverted radio spectrum and it is responsible for nearly 75 per cent of the total emission. A weak extension towards PA$\sim$70 deg is reported by \citet{Patnaik1992} with VLA in A configuration at 8.4 GHz. At higher resolutions, VLBI studies revealed a kpc and a pc-scale jet-like feature in the north-west direction \citep[e.g.][]{Taylor1996,Lister2018} with a flat spectral index \citep[see also][]{Konar2004}. Recently, it has been studied by our group \citep{Bruni2019} finding that the core exhibit an inverted spectrum in the GHz range, which have been interpreted as signature of a possible high-frequency peak, and thus a young age for the radio core, probably restarted. Our observations reveal a compact morphology, either slightly resolved (22 GHz) or unresolved (45 GHz), in agreement with previous studies at similar resolution \citep[][at 15 GHz with VLA in C-configuration]{SaikiaShastriSinhaKapahiSwarup1984}, with flux densities of the order of $\sim$1 mJy. The spectral index is compatible with flat/inverted, in agreement with previous studies. We can interpret this evidence, together with the enhanced radio flux, as due to a jet closely aligned with the line of sight, as suggested by VLBI monitoring of the pc-scale jet in the MOJAVE project \citep[e.g.][]{Lister2019}. However, considering the lower frequencies SED points in \citet{Bruni2019}, this evidence could suggest that the turnover of the spectrum is occurring at these frequencies, confirming the high-frequency peaked scenario.

\smallskip
\noindent\textbf{QSOB0241+62} 
It has been the subject of many VLA and VLBI studies, which revealed a flat-spectrum core plus a jet-like feature in the SE direction lobe morphology \citep[][]{Preston1985,Lister1994,Lister2018}, and the morphology is also reported at 22 and 43 GHz \citep[][]{Charlot2010}. In our VLA maps at 22 and 45 GHz we do not observe hint of jet-like structure at either frequencies, although the source is slightly resolved (but still compact) in both cases. The high measured flux densities are in the range 700 - 900 mJy, with a spectral index which is compatible with flat, which can be interpreted in the light of a jet closely aligned with the line-of-sight, \citep[see the MOJAVE project page][]{Lister2019}, in agreement with the Flat-Spectrum Radio Quasar (FSRQ) classification \citet{Hervet2016}.

\smallskip
\noindent\textbf{NGC1275} 
It is a Sy 2 galaxy identified with the well known radio source 3C84, but \citet{Buttiglione2010} re-classify it as a Low Excitation Galaxy (LEG). VLA maps reveal a flat-spectrum core + extended features in the NS direction, the core having flux densities in the range 20.6 - 22 Jy \citep[6 and 20 cm,][]{HU01}. High resolution observations reveal a radio morphology with multiple lobes at different position angles on different scales, from pc to tens kpc scale \citep[e.g.][]{Pedlar1990,Walker2000,Giovannini2018}. Monitoring studies of 3C84 have revealed that it is characterised by recurrent jet activity in its central regions \citep[e.g.][]{Lister2019}, confirmed at high frequencies \citep[see][]{Suzuki2012}. Our maps reveal a compact morphology, either slightly resolved (22 GHz) or unresolved (45 GHz), with flux densities in the range 17.4 - 23 Jy, and a $\sim$ flat spectral index. Our observations can be explained in terms of a jet closely aligned with the line-of-sight.

\smallskip
\noindent\textbf{NGC788} 
\citet{UlvestadWilson1989} studied it with VLA at 6 and 20 cm ($\sim$ 1 arcsec resolution), reporting an unresolved morphology, fluxes of 1-2 \mjy{} and a borderline spectral index between steep and flat ($\alpha\,\sim$0.47), see also \citet{NagarWilsonMulchaeyGallimore1999}. With 1" 22 GHz observation, \citet{Smith2020} observe an slightly extended morphology, interpreted as due to star formation, with flux density of $\sim$0.61 \mjy{}. Our $\sim$1 arcsec resolution maps reveal a compact, core-dominated morphology, with a 22 GHz flux density of 0.87$\pm$0.06, higher respect to \citet{Smith2020}, and a spectral index of $\alpha\,=\,$0.4$\pm$0.2, compatible with that derived by Ulvestad \& Wilson at similar resolution. 

\smallskip
\noindent\textbf{IGRJ00333+6122}
The only existing work comprising this source is the 1.4 GHz NVSS characterisation made by our group in \citet{Panessa2015}. The flux density at 1.4 GHz is $\sim$9.5 mJy with a compact, unresolved morphology at the 45 arcsec resolution. Even at the $\sim$1 arcsec resolution level of our map the source remain unresolved, with flux densities smaller that $\sim$690 \mujy, with a steep spectral index $\alpha\,\sim$1.25$\pm$0.4 .

\smallskip
\noindent\textbf{LEDA168563} 
The only existing work is by our group \citep{Panessa2015}, who reported flux density of $\sim\,$15 \mjy at 1.4 GHz with NVSS. Our radio maps reveal a compact morphology either slightly resolved (22 GHz) or unresolved (45 GHz), with flux densities in the range 1.5 - 3.2 \mjy, and a steep ($\alpha\,\sim$1.07$\pm$0.12) spectral index.

\smallskip
\noindent\textbf{NGC4151} 
It has been the subject of many studies at different resolution and frequency bands. VLA 5 GHz 1 arcsec maps reveal an elongated structure (P.A. $\sim$84 deg) with a steep ($\sim\,$0.83) spectral index \citep[e.g.][]{UlvestadWilsonSramek1981,WilsonUlvestad1982}. Higher resolution observations have resolved the elongated structure into five \citep[][0.6" resolution]{Johnston1982} and 6 components, with one of them (C4 in their nomenclature) which has a flat spectrum and high brightness temperature, likely to be the AGN nucleus \citep[VLA 0.2" resolution at 15 GHz][]{CarralTurnerHo1990}. Combining VLA and MERLIN observations at 5 and 8 GHz \citet{Pedlar1993} find a two-sided jet at PA$\sim\,$77 deg, confirming that there is a flat-spectrum component exhibiting a 150-mas scale jet. \citet{Mundell2003} with VLA and VLBA observations further resolve the nucleus in three components, see also \citet{Ulvestad2005}. Recently \citet{Williams2017} report a significant variability in the core flux density (factor of 2) in their eMERLIN maps at 1.51 GHz. Our radio maps suggest an elongated morphology of the radio emission at both frequencies at position angles of 75.7$\pm$0.3 and 72$\pm$1 deg at 22 and 45 GHz, respectively, with a spectral index which is steep ($\alpha\,=$0.7$\pm$0.1), in agreement with works at similar resolutions \citep[e.g.][]{BoolerPedlarDavies1982,Johnston1982} and with a jet scenario. 

\smallskip
\noindent\textbf{NGC5252} 
It is a Sy1.9 galaxy which is known to exhibit pair of ionisation cones in optical emission lines images \citep[e.g.][]{TadhunterTsvetanov1989}. Previous VLA studies revealed two components $\sim$10 kpc far, one (the southern) coincident with optical nucleus, the other (northern) which can be associated with emission line features, both having steep spectra \citep[e.g.][]{WilsonTsvetanov1994,Tsvetanov1996,Kukula1995}. \citet{YangYangParagi2017} performed EVN observations focusing on the off-nuclear component, finding that it is compact on pc scale, it has a flat spectrum and a high brightness temperature ($T_{B}\,\ge\,$10$^7$ K), which is interpreted as signature of a background LLAGN. Our radio maps exhibit a compact, resolved core and a northern component $\sim$22 arcsec far ($\sim$ 10 kpc) which is however detected only at 22 GHz. The flux densities for the nuclear component are $\sim$8.8 and 2.7 \mjy{} at 22 and 45 GHz, while the 22 GHz flux of the northern component is $\sim$0.6 \mjy. The nuclear component has a steep spectrum.

\smallskip
\noindent\textbf{Mkn6} 
Previous VLA studies reveal that Mkn 6 possesses radio emission on three spatial scales: i) $\sim$1 kpc scale jet NE-SW direction, with two components $\sim\,$1 arcsec apart; ii) bubbles on scales $\ge$1.5 kpc E-W direction; iii) bubbles on scales $\ge$4.5 kpc \citep[][]{Kharb2006}. This complex structure may be due to either a jet precession and/or the jet interaction with ambient interstellar medium. \citet{Kharb2014} performed VLBA observations spotting the presence of a core with a high brightness temperature ($T_B\,\sim$10$^8$ K), interpreted as the optically-thick base of the jet, the other components of the jets being elongated and having steeper spectra. Our radio maps reveal a resolved morphology elongated in the N-S direction. The double structure seen by e.g. Ulvestad et al. is seen only in the full-resolution Q -band maps, with a separation of $\sim$1 arcsec. Considering the 22 GHz and the naturally-weighted, untapered Q-band maps, the flux densities are of 9.7 and 16.6 \mjy and the spectral index is steep, compatible with previous studies at similar resolution \citep[e.g.][]{vanderHulstCraneKeel981}.

\smallskip
\noindent\textbf{MCG+08-11-11} 


VLA maps at 8.4 GHz revealed a central component (P.A. $\sim$ 128 deg) and an extended one on scales $\sim$1.2 kpc in the N-S direction \citep[][]{WilsonWillis1980,Schmitt2001} while VLBI observations a 2 and 6 cm resolve the center into three components (0.7" length NW-SE direction, P.A. = 127 deg)with a component having a flatter spectrum (recognized as the core) and a 2.4 kpc emission on N-S direction. \citet{Behar2018} observed it at 100 GHz with CARMA finding a flux density of 7.53$\pm$0.19 and a spectral slope, calculated within the CARMA 100 GHz band, of 0.47$\pm$0.29. \citet{Smith2020} 1" resolution 22 GHz map reveal a morphology compatible with ours, with a flux density 0f $\sim$13.8 \mjy{} smaller than our 15.6$\pm$0.8\mjy{}. Our radio maps exhibit a en elongated structure in the N-S direction with P.A. = 160$\pm$1 and 131$\pm$16 deg at 22 and 45 GHz, respectively. The flux densities for the compact component are of order of 15.7 - 9.3 \mjy, with a spectral index of 0.75$\pm$0.11. The above results are compatible with previous VLA works at similar resolutions \citep[e.g.][]{WilsonWillis1980,UlvestadWilson1986}.

\smallskip
\noindent\textbf{Mkn3} 
Previous VLA maps reveal an asymmetric double morphology extending for $\sim$600 pc in the E-W direction, with the two components having steep spectra \citep[e.g.][]{WilsonPooleyWillisClements1980}. High resolutions observations resolve it into eight components in an S-shaped structure, with the closest component to the optical position having flat spectrum and recognized as the nucleus \citep[e.g.][]{Kukula1993,Kukula1999}. Our radio maps reveal a double morphology in the E-W direction, with components which are 1.5 arcsec apart ($\sim$400 pc). Our maps reveal a double morphology in the E-W direction, on a scale of $\sim$1 kpc, in which both the components have steep spectra. The low (0.6) concentration index of the core (East) component suggests that additional optically-thin, steep spectrum jet components may hide in the kpc-component and significantly contribute at these frequencies.

\smallskip
\noindent\textbf{NGC1142} 
It is part of the system \textit{Arp118}, comprising also the spheroidal galaxy NGC1143 (separation $\sim$40 arcsec), and it is most probably the result of a collisional encounter \citep[e.g.][]{JoyGhigo1988}. The system presents a morphology with a southern ridge of radio emission, corresponding to optical emission line knots, a compact component close to the optical position and a group of three compact components (further resolved into 6) $\sim$8 arcsec far in the NE direction \citep[e.g.][]{JoyGhigo1988,Condon1990}, both having flat spectra \citep{Thean2000}. The triple system has been further by \citet{Thean2001} into 6 components with MERLIN observations. Our radio maps reveal a complex radio morphology, in which the southern ring of radio emission is visible in the 22 GHz maps, but not in the 45 GHz map. We detect 6 components in the 22 GHz map, but only three are detected also in the naturally-weighted, tapered 45 GHz maps. The morphology of radio emission is similar to previous works at similar resolution \citep[e.g.][]{Thean2000}, but we detect (with high significance) an additional component with respect to previous VLA studies, i.e. component C, only detected at 22 GHz. Only the components A1, A2 and B (where the nucleus is believed to reside) are detected at both bands. The spectral indices for A1 and A2 are steep, while the spectral index for component B is flatter (although borderline, $\alpha\,\sim\,=$0.5$\pm$0.13). The separation between the core component and the A1/2 double is $\sim$8 arcsec, corresponding to $\sim$4.5 kpc.

\smallskip
\noindent\textbf{NGC4388} 
It is a Sy2 edge-on galaxy in the \textit{Virgo} cluster, and it is known for having wide ionisation cones extending well above and below the stellar disk \citep[e.g.][]{AyaniIye1989}. VLA observations reveal a cross-like morphology extending for in the N direction $\sim$19 arcsec (P.A. $\sim$10 deg), with a central double of $\sim$1.9 arcsec, constituted by a southern component which may be associated to a spike in optical emission lines \citep[e.g.][]{HummelSaikia1991,FalckeWilsonSimpson1998,HU01,DamasSegovia2016} and a northern component forming a collimated jet \citep[e.g.][]{Kukula1999,Mundell2000}. The nucleus is detected by \citet{GirolettiPanessa2009} with EVN only at 1.6 GHz, with a $T_B\,\sim$10$^6$ K, which can not rule out a thermal origin. \citet{Smith2016} studied it with VLA in C configuration at 22 GHz ($\sim$1 arcsec resolution), finding an elongated morphology, but no double component, with flux density $\sim$3.26 \mjy. \citet{Behar2018} observed it with CARMA at 100 GHz reporting flux density of 3.58$\pm$0.34 \mjy. Our radio maps reveal a double morphology in the direction NE-SW, with a ridge of radio emission between the components which are $\sim$1.9 arcsec apart ($\sim$330 pc). The spectral slope of the northern component (where the nucleus is believed to reside) is flat ($\alpha\,=$0.27$\pm$0.11) while the southern component exhibits a steeper spectrum ($\alpha\,=$0.87$\pm$0.09). The position quoted by \citet{Smith2016} is compatible with our northern component, and the flux density of our northern component is compatible with both their 22 GHz flux and the 100 GHz flux of \citet{Behar2018}, the same being valid for the high-frequency spectral slope.

\smallskip
\noindent\textbf{NGC1068} 
It is a well studied Seyfert 2 galaxy, which is know to have bipolar NLR outflowing structures apparently aligned with the radio jet \citep[e.g.][]{CrenshawKraemer2000,Whittle2005}. We keep the nomenclature of \citet{WilsonUlvestad1983}, for previous VLA and VLBI maps see \citep[e.g.][]{WilsonWillis1980,Wilson1981,vanderHulstHummelDichey1982,WilsonUlvestad1982,KukulaPedlarBaumODea1995}. The \citet{WilsonUlvestad1983} maps at 4.9 and 15 GHz ($\sim$0.4 arcsec resolution) reveal four components aligned in the NE-SW direction on a scale of $\sim$13 arcsec, interpreted as a radio jet emanating from the nucleus, which is resolved into three structure. The steep-spectrum NE lobe is find to be coincident with the edge of the optical line emitting clouds. High resolution observations \citep[e.g.][]{PedlarBoolerSpencerStewart1983,UlvestadNeffWilson1987,MuxlowPedlarHollowayGallimoreAntonucci1996} resolved the central component into several components constituting a jet. However, in a series of paper, Gallimore et al. performed high resolution observations on the central component of the source concluding that the component having the flattest spectrum has a low $T_{B}\,\sim\,3\,\times\,$10$^5$ K brightness temperature, which would favour a thermal bremsstrahlung emission from inner edge of a torus instead of synchrotron self-absorption from the base of a jet \citep[e.g.][]{Gallimore1996Feb,GAllimore1996June,Gallimore2004}, but see also \citet{RoyColbertWilsonUlvestad1998,CottonJaffePerrinWoillez2008}. Our radio maps reveal a morphology compatible with previous VLA observations at comparable resolution, with an overall extension in the SE-NW direction of $\sim$16 arcsec ($\sim$1.3 kpc). The (a) component presents flux densities of 151 - 83 \mjy and a steep spectral index ($\alpha\,=$+0.88$\pm$0.11), in agreement with previous works.
The component (b), where high-resolution observations have identified the nucleus, the flux densities are in the range 129 - 72 \mjy with a spectral index $\alpha\,=$+0.85$\pm$0.10, again compatible with previous works at similar resolution. The component (c) in \citet{WilsonUlvestad1983} is only seen at 22 GHz with flux density TOT \mjy. The extended component in the SW \citep[not considered in][]{WilsonUlvestad1983} exhibit a flatter spectrum ($\alpha\,=$0.47$\pm$0.14) and flux densities of 16 - 12 \mjy{} .

\begin{figure*}\scriptsize
\centering
\includegraphics[scale=0.75]{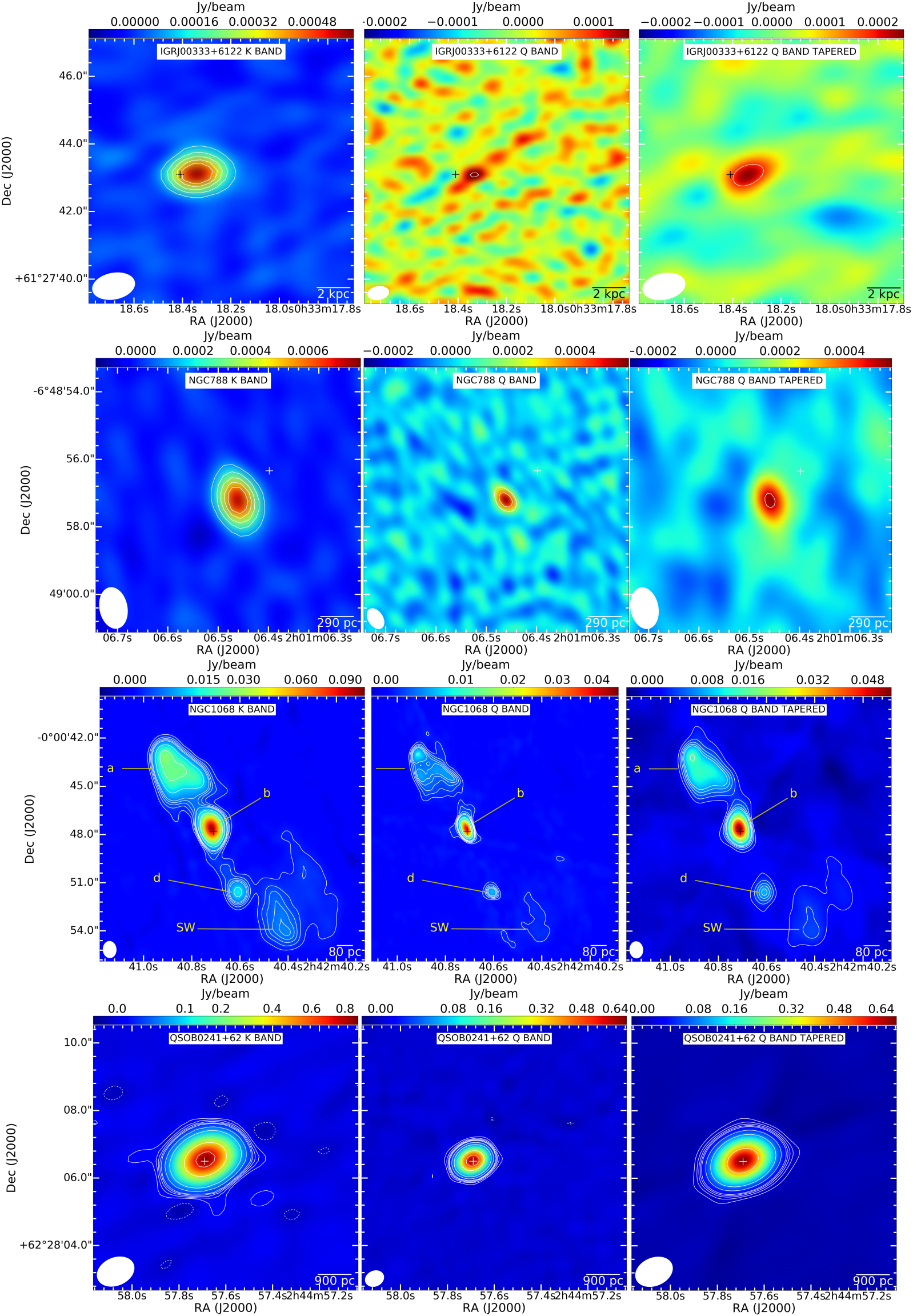}
\caption{Contour and coloured maps for IGRJ00333+6122, NGC788, NGC1068 and QSOB0241+62 (from top to bottom). For each source we show the map at the K band (left), at Q band (centre) and at Q band after tapering and natural weighting (right). The contours are shown at the [-5,5,10,15,20,25,50,100,500,1000,5000] $\times$ $\sigma_{image}$, using the image RMS reported in Table \ref{table:fluxTable}. The horizontal bar in the bottom-right corner of each panel represents the linear scale corresponding to 1 arcsec. The cross corresponds to the position quoted in Table \ref{table:fluxTable}.}
\label{fig:Mappe1}
\end{figure*}

\begin{figure*}\scriptsize
\centering
\includegraphics[scale=0.75]{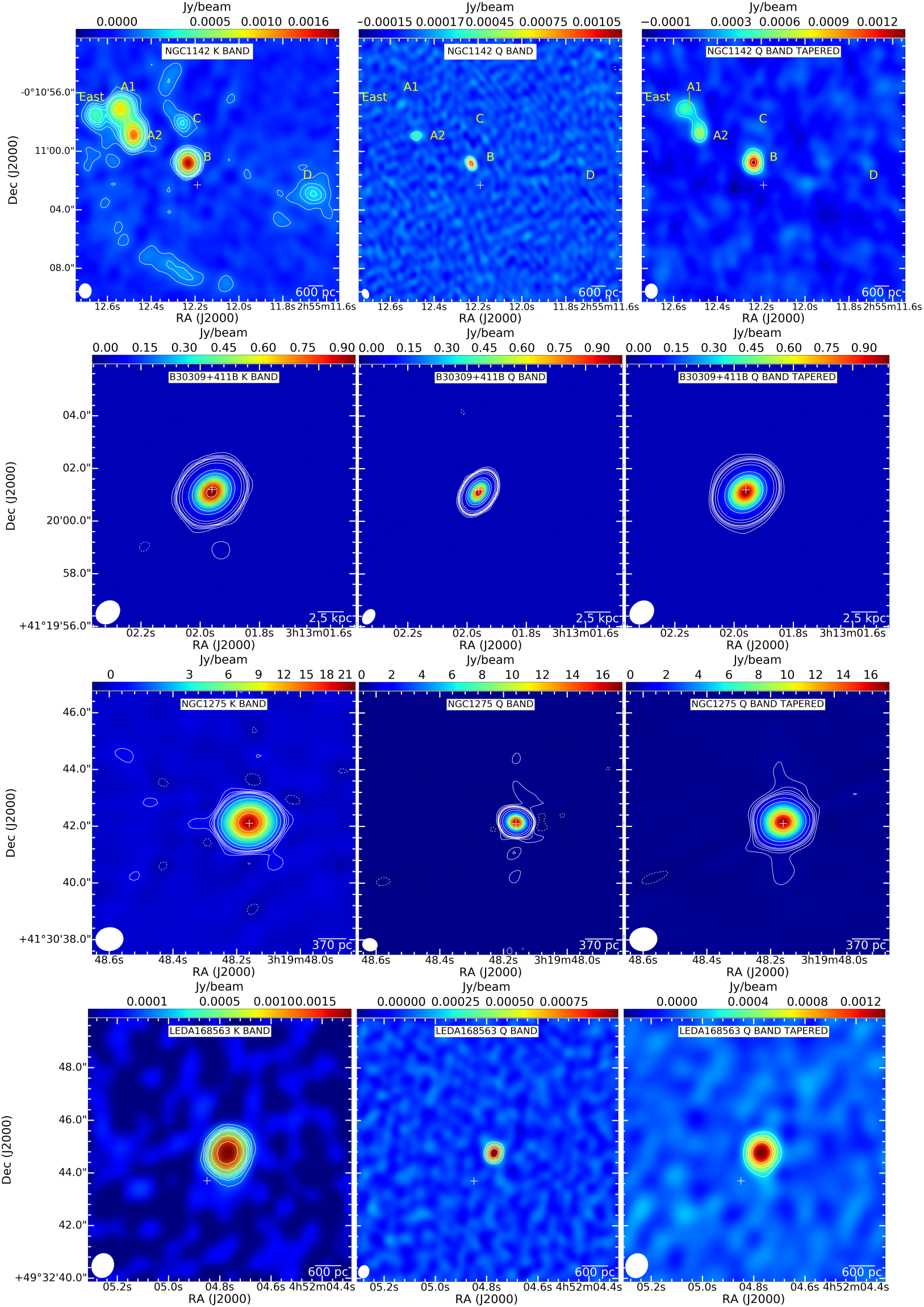}
\caption{Contour and coloured maps for NGC1142, B30309+411B, NGC1275 and LEDA168563 (from top to bottom). For each source we show the map at the K band (left), at Q band (centre) and at Q band after tapering and natural weighting (right). The contours are shown at the [-5,5,10,15,20,25,50,100,500,1000,5000] $\times$ $\sigma_{image}$, using the image RMS reported in table Table \ref{table:fluxTable}.  The horizontal bar in the bottom-right corner of each panel represents the linear scale corresponding to 1 arcsec. The cross corresponds to the position quoted in Table \ref{table:fluxTable}.}
\label{fig:Mappe2}
\end{figure*}

\begin{figure*}\scriptsize
\centering
\includegraphics[scale=0.75]{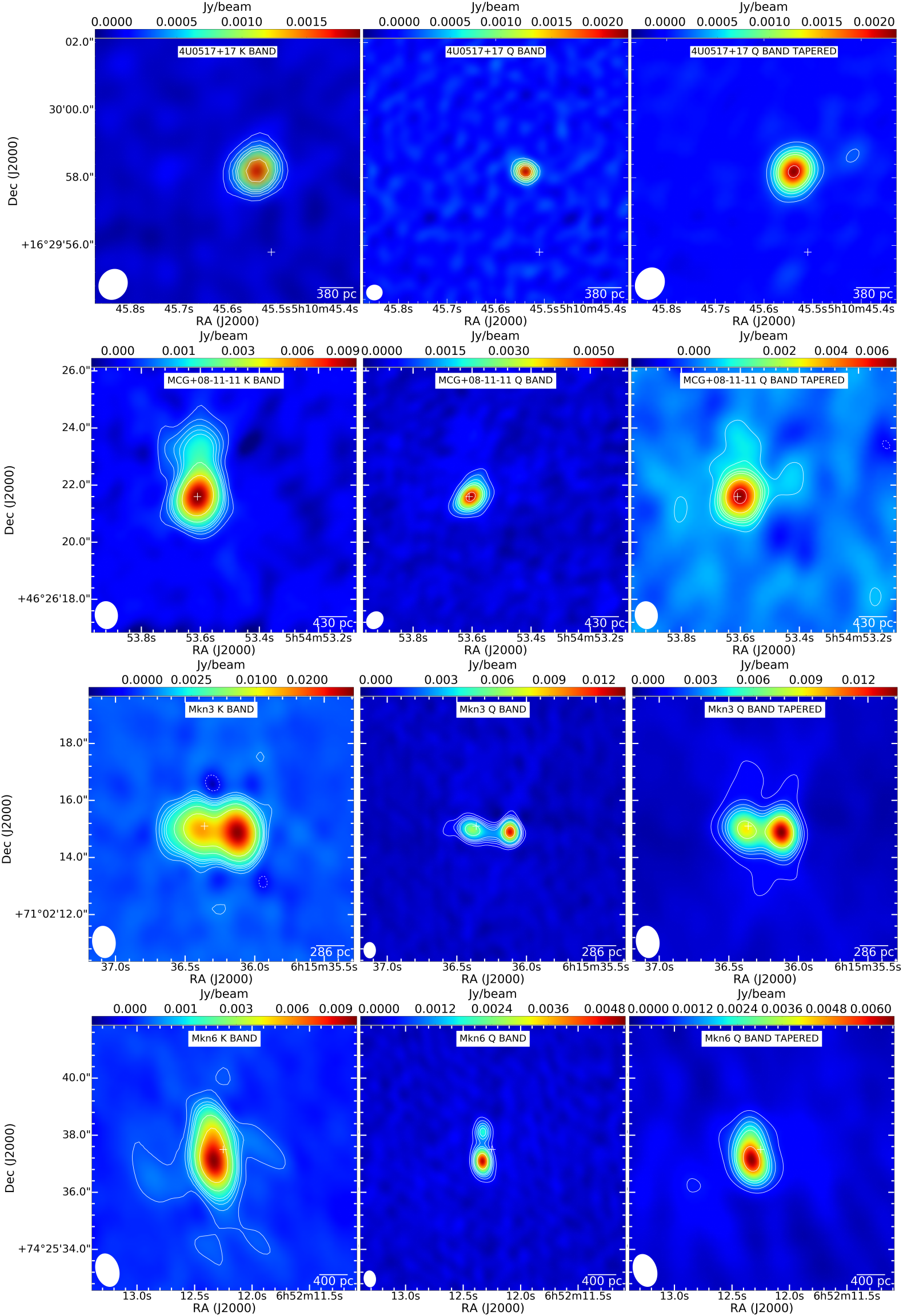}
\caption{Contour and coloured maps for 4U0517+17, MCG+08-11-11, Mkn3 and Mkn6 (from top to bottom). For each source we show the map at the K band (left), at Q band (centre) and at Q band after tapering and natural weighting (right). The contours are shown at the [-5,5,10,15,20,25,50,100,500,1000,5000] $\times$ $\sigma_{image}$, using the image RMS reported in table Table \ref{table:fluxTable}.  The horizontal bar in the bottom-right corner of each panel represents the linear scale corresponding to 1 arcsec.The cross corresponds to the position quoted in Table \ref{table:fluxTable}.}
\label{fig:Mappe3}
\end{figure*}

\begin{figure*}\scriptsize
\centering
\includegraphics[scale=0.8]{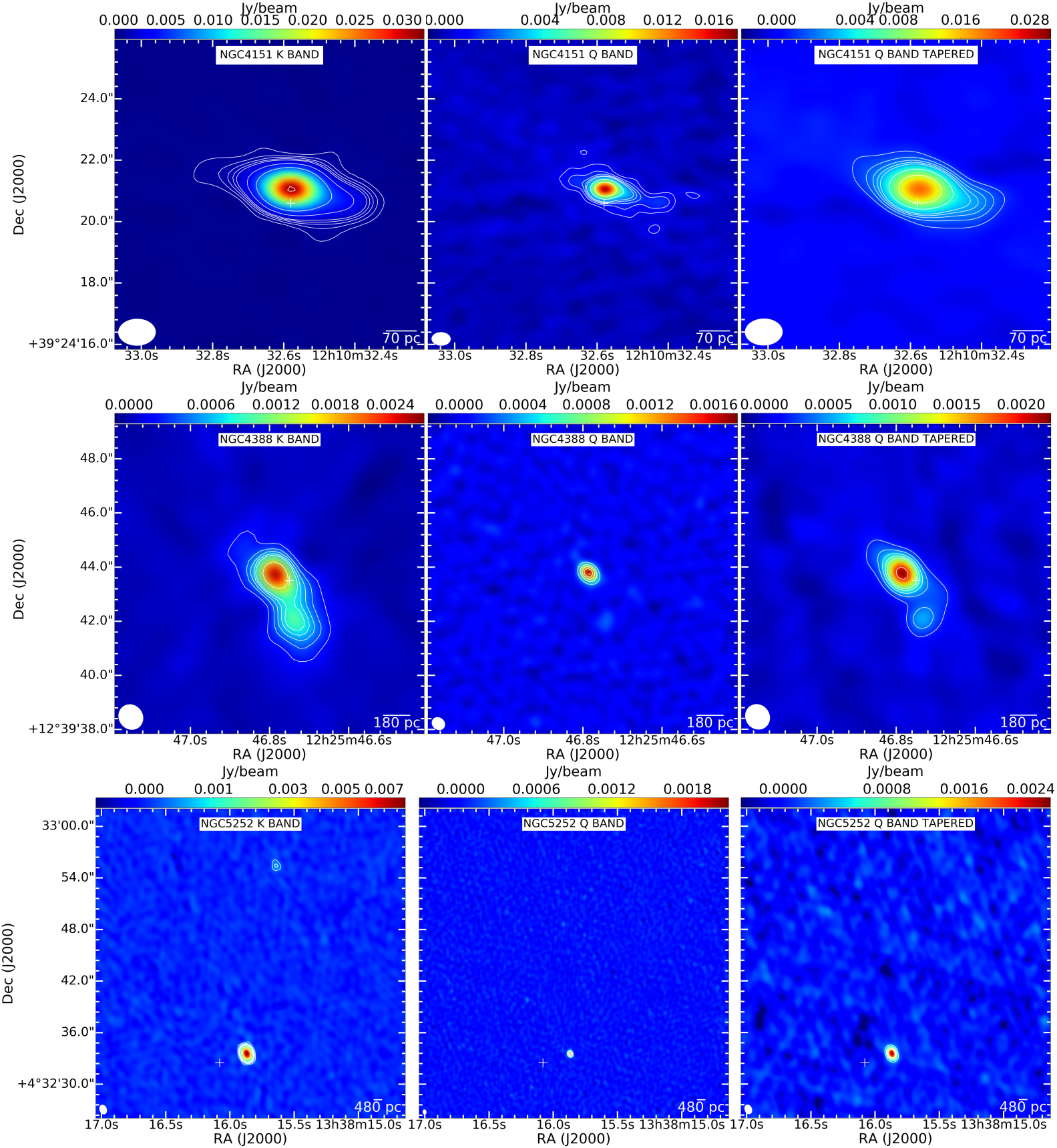}
\caption{Contour and coloured maps for NGC4151, NGC4388 and NGC5252 (from top to bottom). For each source we show the map at the K band (left), at Q band (centre) and at Q band after tapering and natural weighting (right). The contours are shown at the [-5,5,10,15,20,25,50,100,500,1000,5000] $\times$ $\sigma_{image}$, using the image RMS reported in table Table \ref{table:fluxTable}.  The horizontal bar in the bottom-right corner of each panel represents the linear scale corresponding to 1 arcsec. The cross corresponds to the position quoted in Table \ref{table:fluxTable}.}
\label{fig:Mappe4}
\end{figure*}

\bsp	
\label{lastpage}

\onecolumn
\begin{landscape}
\scriptsize
\begin{longtable}{ cccccccccccc }
\caption{Imaging results for the sources in our sample. \textit{Columns:} (1) Target name; (2) Frequency band; (3) Image noise rms [\mujybm], unless specified; (4) Component (only for the resolved sources); (5) Integrated flux density (mJy); (6) Peak intensity (\mjybm); (7) Deconvolved FWHM dimensions (major $\times$ minor axis) for the fitted source, determined from an elliptical Gaussian fit source size (milliarcsec); (8) Source position angle (deg); (9) $\&$ (10) Detected source position in epoch J2000 (hh:mm:ss and \deg:\arcm:\arcs); (11) Spectral index calculated from the 22 GHz and the 45 GHz naturally-weighted, tapered maps. For NGC5252 only the spectral index of the southern (core) component is shown as the northern component is not associated with the AGN (see following sections).}
\label{table:fluxTable}\\
\hline
Target & Band & $\sigma_{\rm image}$ & Component & F$_{\mathrm{total}}$ & F$_{\mathrm{peak}}$ & $\theta_{\mathrm{M}} \times \theta_{\mathrm{m}}$ & P.A. & $\alpha_{\rm J2000}$ & $\delta_{\rm J2000}$ & $\alpha$ \\ 
(1) & (2) & (3) & (4) & (5) & (6) & (7) & (8) & (9) & (10) & (11)\\ 
\hline
\endhead

NGC788 
 & K & 24 & & 0.87$\pm$0.06  & 0.80$\pm$0.05 & $<$0.6 $\times$ 0.4 & $\cdots$ & 02:01:06.46$\pm$0.01 & -06.48.57.22$\pm$0.02 & +0.4$\pm$0.25\\ 
 & Q & 41 & & 0.62$\pm$0.05  & 0.56$\pm$0.07 & $<$0.3 $\times$ 0.2 & $\cdots$ & 02:01:06.46$\pm$0.015 & -06.48.57.20$\pm$0.02 & \\ 
 & Q-tapered & 44 & & 0.66$\pm$0.10  & 0.50$\pm$0.05 & (930$\pm$350) $\times$ (270$\pm$180) & 180$\pm$30 & 02:01:06.45$\pm$0.03 & -06.48.57.30$\pm$0.07 & \\ 
\hline

4U0517+17 
 & K & 26 & & 2.08$\pm$0.12  & 1.9$\pm$0.10 & (300$\pm$60) $\times$ (200$\pm$150) & 80$\pm$60 & 05:10:45.540$\pm$0.005 & +16.29.58.214$\pm$0.006 & -0.09$\pm$0.11\\ 
 & Q & 62 & & 2.42$\pm$0.16  & 2.31$\pm$0.13 & $<$0.2 $\times$ 0.2 & $\cdots$ & 05:10:45.539$\pm$0.006 & +16.29.58.182$\pm$0.005 & \\ 
 & Q-tapered & 38 & & 2.23$\pm$0.13  & 2.19$\pm$0.13 & $<$0.5 $\times$ 0.4 & $\cdots$ & 05:10:45.538$\pm$0.006 & +16.29.58.180$\pm$0.007 & \\ 
\hline

IGRJ00333+6122
 & K & 17 & & 0.70$\pm$0.05  & 0.62$\pm$0.04 & $<$0.6 $\times$ 0.4 & $\cdots$ & 00:33:18.34$\pm$0.02 & +61.27.43.13$\pm$0.01 & +1.2$\pm$0.4\\ 
 & Q & 38 & & 0.32$\pm$0.09  & 0.17$\pm$0.03 & $<$0.3 $\times$ 0.2 & $\cdots$ & 00:33:18.33$\pm$0.13 & +61.27.43.1$\pm$0.1 & \\ 
 & Q-tapered & 36 & & 0.29$\pm$0.07  & 0.25$\pm$0.07 & $<$0.6 $\times$ 0.4 & $\cdots$ & 00:33:18.33$\pm$0.09 & +61.27.43.11$\pm$0.05 & \\ 
\hline

B30309+411B
 & K & 181 & & 1.05$\pm$0.05 Jy & 1.05$\pm$0.05 Jy/beam & (45$\pm$4) $\times$ (30$\pm$10) & 19$\pm$17 & 03:13:01.960$\pm$0.005 & +41.20.01.100$\pm$0.004 & +0.06$\pm$0.1\\ 
 & Q & 157 & & 1.01$\pm$0.05 Jy & 1.00$\pm$0.05 Jy/beam & (52$\pm$2) $\times$ (28$\pm$2) & 130$\pm$3 & 03:13:01.961$\pm$0.004 & +41.20.01.097$\pm$0.006 & \\ 
 & Q-tapered & 260 & & 1.01$\pm$0.05  Jy & 1.01$\pm$0.05 Jy/beam & $<$0.5 $\times$ 0.4 & $\cdots$ & 03:13:01.960$\pm$0.006 & +41.20.01.097$\pm$0.005 & \\ 
\hline

LEDA168563
 & K & 17 & & 3.19$\pm$0.16  & 3.0$\pm$0.15 & (245$\pm$30) $\times$ (190$\pm$40) & $\cdots$ & 04:52:04.770$\pm$0.004 & +49.32.44.759$\pm$0.005 & +1.07$\pm$0.12\\ 
 & Q & 40 & & 1.6$\pm$0.1  & 1.45$\pm$0.08 & (160$\pm$50) $\times$ (50$\pm$90) & $\cdots$ & 04:52:04.771$\pm$0.005 & +49.32.44.764$\pm$0.006 & \\ 
 & Q-tapered & 38 & & 1.6$\pm$0.1  & 1.56$\pm$0.09 & $<$0.5 $\times$ 0.4 & $\cdots$ & 04:52:04.770$\pm$0.005 & +49.32.44.767$\pm$0.007 & \\ 
\hline

NGC4151
 & K & 63 & & 43.5$\pm$2.2  & 31.0$\pm$1.5 & (1060$\pm$6) $\times$ (258$\pm$7) & 76.0$\pm$0.3 & 12:10:32.575$\pm$0.005 & +39.24.21.050$\pm$0.006 & +0.7$\pm$0.1\\ 
 & Q & 86 & & 22.6$\pm$1.15  & 16.2$\pm$0.8 & (500$\pm$9) $\times$ (174$\pm$7) & 78$\pm$1 & 12:10:32.575$\pm$0.004 & +39.24.21.047$\pm$0.006 & \\
 & Q - tapered & 102 & & 26.2$\pm$1.3  & 19.4$\pm$0.9 & (980$\pm$20) $\times$ (240$\pm$20) & 72$\pm$2 & 12:10:32.572$\pm$0.006 & +39.24.21.041$\pm$0.005 & \\
\hline

IGRJ16426+6536
 & K & 8 & $\cdots$  & $<$0.024 & $\cdots$ & $\cdots$ & $\cdots$ & $\cdots$\\
 & Q & 54 & $\cdots$  & $<$0.162 & $\cdots$ & $\cdots$ & $\cdots$ & $\cdots$\\
\hline

QSOB0241+62
 & K & 740 & & 902$\pm$45 & 895$\pm$45 & (102$\pm$13) $\times$ (50$\pm$30) & 63$\pm$15 & 02:44:57.687$\pm$0.004 & +62.28.06.549$\pm$0.005 & +0.3$\pm$0.1\\ 
 & Q & 505 & & 706$\pm$35 & 702$\pm$35 & (48$\pm$6) $\times$ (13$\pm$16) & 175$\pm$12 & 02:44:57.692$\pm$0.005 & +62.28.06.513$\pm$0.004 & \\ 
 & Q - tapered & 872 & & 706$\pm$35 & 701$\pm$35 & (100$\pm$20) $\times$ (50$\pm$30) & 60$\pm$30 & 02:44:57.690$\pm$0.005 & +62.28.06.518$\pm$0.006 & \\ 
 \hline
 
 NGC1275
 & K & 14 \mjybm{} & & 23.0$\pm$1.1 Jy & 22.9$\pm$1.1 Jy beam$^{-1}$ & (80$\pm$10) $\times$ (52$\pm$20) & 57$\pm$20 & 03:19:48.161$\pm$0.004 & +41.30.42.131$\pm$0.005 & +0.4$\pm$0.1\\ 
 & Q & 17 \mjybm{} & & 17.7$\pm$0.9 Jy & 17.7$\pm$0.9 Jy beam$^{-1}$ & $<$0.3 $\times$ 0.2 & $\cdots$ & 03:19:48.160$\pm$0.007 & +41.30.42.140$\pm$0.006 & \\
 & Q - tapered & 25 \mjybm{} & & 17.5$\pm$0.9 Jy & 17.4$\pm$0.9 Jy beam$^{-1}$ & $<$0.5 $\times$ 0.4 & $\cdots$ & 03:19:48.161$\pm$0.006 & +41.30.42.153$\pm$0.007 & \\ 
 \hline
 
 Mkn6
 & K & 46 & & 16.6$\pm$0.8 & 10.3$\pm$0.5 & (1360$\pm$14) $\times$ (250$\pm$14) & 1.9$\pm$0.3 & 06:52:12.331$\pm$0.005 & +74.25.37.245$\pm$0.004 & +0.8$\pm$0.1\\
 & Q (south)& 64 & & 6.7$\pm$0.4 & 5.0$\pm$0.3 & (440$\pm$20)$\times$(160$\pm$15) & 5$\pm$2 & 06:52:12.330$\pm$0.006 & +74.25.37.086$\pm$0.005\\
& Q (north) & & & 2.25$\pm$0.2 & 1.9$\pm$0.1 & (330$\pm$65)$\times$(140$\pm$105) & 170$\pm$20 & 06:52:12.328$\pm$0.005 & +74.25.38.09$\pm$0.01\\
 & Q-tapered & 100 & & 9.7$\pm$0.5  & 6.2$\pm$0.3 & (1000$\pm$50) $\times$ (500$\pm$40) & $\cdots$ & 06:52:12.329$\pm$0.006 & +74.25.37.24$\pm$0.01\\
 \hline
 
  NGC5252
 & K & 41 & Core & 8.8$\pm$0.4  & 8.15$\pm$0.4 & (310$\pm$30) $\times$ (225$\pm$25) & $\cdots$ & 13:38:15.869$\pm$0.004 & +04.32.33.54$\pm$0.02 & +1.7$\pm$0.1\\ 
 & K & & North & 0.6$\pm$0.1  & 0.44$\pm$0.04 & (850$\pm$350) $\times$ (245$\pm$210) & $\cdots$ & 13:38:15.64$\pm$0.03 & +04.32.55.42$\pm$0.07\\
 & Q & 100 & Core & 2.35$\pm$0.2  & 2.2$\pm$0.15 & (164$\pm$100) $\times$ (50$\pm$110) & $\cdots$ & 13:38:15.868$\pm$0.008 & +04.32.33.52$\pm$0.02\\ 
 & Q & & North & $\cdots$  & $<$0.3 & $\cdots$ & $\cdots$ & $\cdots$ & $\cdots$\\
 & Q - tapered & 62 & Core & 2.7$\pm$0.2  & 2.5$\pm$0.14 & (370$\pm$160) $\times$ (240$\pm$130) & $\cdots$ & 13:38:15.871$\pm$0.009 & +04.32.33.55$\pm$0.01\\ 
 & Q - tapered & & North & $\cdots$  & $<$0.2 & $\cdots$ & $\cdots$ & $\cdots$ & $\cdots$\\ 
 \hline
 
 Mkn3 
 & K & 80 & W & 34$\pm$4  & 31.7$\pm$1.6 & (275$\pm$14) $\times$ (220$\pm$20) & 130$\pm$13 & 06:15:36.120$\pm$0.007 & +71.02.14.875$\pm$0.005 & +1.2$\pm$0.1\\
 &  & & E & 25.7$\pm$0.8 & 15.3$\pm$0.8 & (1080$\pm$13) $\times$ (250$\pm$40)  & 70$\pm$1 & 06:15:36.382$\pm$0.005 & +71.02.14.947$\pm$0.006 & +0.87$\pm$0.09\\
 & Q & 86 & W & 14.6$\pm$0.7 & 12.5$\pm$0.7 & (243$\pm$9) $\times$ (90$\pm$44) & 99$\pm$5 & 06:15:36.119$\pm$0.005 & +71.02.14.894$\pm$0.004\\
 &  & & E & 12.7$\pm$0.7  & 6.0$\pm$0.3 & $<$0.3 $\times$ $<$0.2  & $\cdots$ & 06:15:36.375$\pm$0.007 & +71.02.14.956$\pm$0.006\\
 & Q-tapered & 150 & W & 15.1$\pm$0.8  & 13.6$\pm$0.7 & (424$\pm$50) $\times$ (210$\pm$70) & 170$\pm$12 & 06:15:36.117$\pm$0.006 & +71.02.14.883$\pm$0.007\\
 &  & & E & 14.1$\pm$0.8  & 8.6$\pm$0.5 & (900$\pm$55) $\times$ (560$\pm$80)  & 67$\pm$8 & 06:15:36.375$\pm$0.009 & +71.02.14.97$\pm$0.01\\
\hline

NGC4388
 & K & 32 & NE & 3.5$\pm$0.2  & 2.6$\pm$0.1 & (700$\pm$30) $\times$ (370$\pm$35) & 52$\pm$4 & 12:25:46.786$\pm$0.006 & +12.39.43.732$\pm$0.006 & +0.27$\pm$0.11\\
 &  & & SW & 3.1$\pm$0.2  & 0.9$\pm$0.05 & (1825$\pm$95) $\times$ (1080$\pm$65) & 13$\pm$4 & 12:25:46.73$\pm$0.02 & +12.39.42.13$\pm$0.03 & +1.6$\pm$0.26\\
 & Q & 67 & NE & 2.16$\pm$0.2  & 1.8$\pm$0.1 & (310$\pm$60) $\times$ (100$\pm$60) & 40$\pm$16 & 12:25:46.785$\pm$0.009 & +12.39.43.774$\pm$0.009\\
 &  & & SW & $\cdots$  & $\cdots$ & $\cdots$ & $\cdots$ & $\cdots$ & $\cdots$\\
 & Q-tapered & 40 & NE & 2.9$\pm$0.2  & 2.14$\pm$0.1 & (784$\pm$45) $\times$ (270$\pm$70) & 49$\pm$4 & 12:25:46.786$\pm$0.009 & +12.39.43.772$\pm$0.006\\
 &  & & SW & 1.03$\pm$0.14  & 0.50$\pm$0.05 & (1520$\pm$20) $\times$ (480$\pm$24) & 154$\pm$6 & 12:25:46.72$\pm$0.05 & +12.39.42.22$\pm$0.07\\
\hline

NGC1142
 & K & 10 & East & 0.87$\pm$0.04  & 0.34$\pm$0.01 & (1500$\pm$70) $\times$ (825$\pm$65) & 100$\pm$4 & 02:55:12.63$\pm$0.02 & -00.10.57.526$\pm$0.02\\
 & & & A1 & 2.0$\pm$0.1  & 0.93$\pm$0.05 & (1050$\pm$30) $\times$ (940$\pm$30) & 42$\pm$14 & 02:55:12.539$\pm$0.006 & -00.10.57.182$\pm$0.007 & +0.9$\pm$0.25\\
 &  & & A2 & 2.25$\pm$0.1  & 1.32$\pm$0.07 & (1060$\pm$20) $\times$ (540$\pm$30) & 13$\pm$1 & 02:55:12.482$\pm$0.005 & -00.10.58.806$\pm$0.006 & +1.4$\pm$0.3\\
 &  & & B & 2.24$\pm$0.1  & 2.1$\pm$0.1 & (300$\pm$15) $\times$ (60$\pm$60) & 84$\pm$6 & 02:55:12.232$\pm$0.001 & -00.11.00.804$\pm$0.005 & +0.5$\pm$0.13\\
 &  & & D & 0.71$\pm$0.05  & 0.18$\pm$0.01 & (1780$\pm$140) $\times$ (1440$\pm$130) & 85$\pm$20 & 02:55:11.66$\pm$0.05 & -00.11.02.89$\pm$0.04\\
  & Q & 42 & East & $\cdots$  & $<$0.12 & $\cdots$ & $\cdots$ & $\cdots$ & $\cdots$\\
 & & & A1 & 0.17$\pm$0.06  & 0.21$\pm$0.04 & $<$0.3 $\times$ $<$0.2 & $\cdots$ & 02:55:12.55$\pm$0.05 & -00.10.56.79$\pm$0.05\\
 &  & & A2 & 0.5$\pm$0.1  & 0.41$\pm$0.05 & $<$0.3 $\times$ $<$0.2 & $\cdots$ & 02:55:12.480$\pm$0.004 & -00.10.58.93$\pm$0.02\\
 &  & & B & 1.2$\pm$0.1  & 1.16$\pm$0.07 & $<$0.3 $\times$ $<$0.2 & $\cdots$ & 02:55:12.231$\pm$0.007 & -00.11.00.82$\pm$0.01\\
 &  & & D & $\cdots$  & $<$0.12 & $\cdots$ & $\cdots$ & $\cdots$ & $\cdots$\\
 & Q-tapered & 51 & East & $\cdots$  & $<$0.15 & $\cdots$ & $\cdots$ & $\cdots$ & $\cdots$\\
 & & & A1 & 1.0$\pm$0.2  & 0.45$\pm$0.06 & (1360$\pm$30) $\times$ (780$\pm$30) & 70$\pm$20 & 02:55:12.54$\pm$0.08 & -00.10.57.17$\pm$0.06\\
 &  & & A2 & 0.85$\pm$0.2  & 0.67$\pm$0.06 & (1380$\pm$40) $\times$ (490$\pm$40) & 16$\pm$18 & 02:55:12.48$\pm$0.02 & -00.10.58.63$\pm$0.05\\
 &  & & B & 1.6$\pm$0.1 & 1.4$\pm$0.1 & (470$\pm$130) $\times$ (290$\pm$190) & 17$\pm$82 & 02:55:12.24$\pm$0.014 & -00.11.00.72$\pm$0.02\\
 &  & & D & $\cdots$  & $<$0.153 & $\cdots$ & $\cdots$ & $\cdots$ & $\cdots$\\
 \hline
 
 NGC1068
 & K & 208 & a & 151.7$\pm$7.7  & $\cdots$ & 2700$\times$1500 & $\cdots$ & $\cdots$ & $\cdots$ & +0.9$\pm$0.1\\
 &  & & b & 128.8$\pm$6.4  & 98.8$\pm$4.9 & (767$\pm$6) $\times$ (252$\pm$8) & 23.0$\pm$0.5 & 02:42:40.7185$\pm$0.008 & -00.00.47.607$\pm$0.007 & +0.8$\pm$0.1\\
 &  & & d & 16.2$\pm$0.9  & 12.2$\pm$0.7 & (620$\pm$60) $\times$ (460$\pm$60) & 33$\pm$17 & 02:42:40.609$\pm$0.007 & -00.00.51.55$\pm$0.01 & +0.47$\pm$0.01\\
 &  & & SW & 60.8$\pm$3.8  & $\cdots$ & 4200$\times$3100 & $\cdots$ & $\cdots$ & $\cdots$ & +0.14$\pm$0.13\\
 & Q & 160 & a & 88.6$\pm$4.8  & $\cdots$ & 2720$\times$1560 & $\cdots$ & $\cdots$ & $\cdots$\\
 &  & & b & 70.4$\pm$3.5  & 40.4$\pm$2.0 & (733$\pm$6) $\times$ (185$\pm$5) & 19.0$\pm$0.3 & 02:42:40.713$\pm$0.009 & -00.00.47.635$\pm$0.007\\
 &  & & d & 9.2$\pm$0.6  & 6.0$\pm$0.4 & (420$\pm$50) $\times$ (308$\pm$60) & 80$\pm$20 & 02:42:40.607$\pm$0.007 & -00.00.51.568$\pm$0.008\\
 &  & & SW & 57.6$\pm$4.8  & $\cdots$ & 5950$\times$3200 & $\cdots$ & $\cdots$ & $\cdots$\\
 & Q - tapered & 295 & a & 82.7$\pm$4.5  & $\cdots$ & 2600$\times$1500 & $\cdots$ & $\cdots$ & $\cdots$\\
 &  & & b & 71.9$\pm$3.7  & 54.9$\pm$2.8 & (810$\pm$15) $\times$ (220$\pm$20) & 14$\pm$1 & 02:42:40.714$\pm$0.005 & -00.00.47.615$\pm$0.006\\
 &  & & d & 12.0$\pm$1.0 & 7.3$\pm$0.5 & (900$\pm$100) $\times$ (600$\pm$100) & 170$\pm$20 & 02:42:40.612$\pm$0.02 & -00.00.51.58$\pm$0.03\\
 &  & & SW & 55.3$\pm$3.4  & $\cdots$ & 6100$\times$3000 & $\cdots$ & $\cdots$ & $\cdots$\\
 \hline
 
MCG+08-11-11
 & K & 40 & core & 15.6$\pm$0.8  & 10.2$\pm$0.5 & (840$\pm$10) $\times$ (440$\pm$10) & 160$\pm$1 & 05:54:53.607$\pm$0.006 & +46.26.21.617$\pm$0.004 & +0.7$\pm$0.1\\
 &  & & jet & 2.8$\pm$0.8  & $\cdots$ & $\cdots$  & $\cdots$ & $\cdots$ & $\cdots$ & +1.1$\pm$0.4\\
 &  & & total & 18.8$\pm$1.2  & $\cdots$ & $\cdots$  & $\cdots$ & $\cdots$ & $\cdots$ & +0.79$\pm$0.14\\
 & Q & 98 & core & 9.4$\pm$0.5  & 5.4$\pm$0.3 & (70$\pm$30) $\times$ (400$\pm$2) & 140$\pm$4 & 05:54:53.602$\pm$0.007 & +46.26.21.595$\pm$0.007\\
 & Q-tapered & 116 & core & 9.3$\pm$0.5 & 7.1$\pm$0.4 & (570$\pm$50) $\times$ (390$\pm$80) & 130$\pm$20 & 05:54:53.600$\pm$0.006 & +46.26.21.609$\pm$0.008\\
 &  & & jet & 1.3$\pm$0.1  & $\cdots$ & $\cdots$  & $\cdots$ & $\cdots$ & $\cdots$\\
 &  & & total & 10.9$\pm$0.8  & $\cdots$ & $\cdots$  & $\cdots$ & $\cdots$ & $\cdots$\\
\hline
\hline
\end{longtable}
\end{landscape}
\twocolumn




\end{document}